\documentclass[aps,prd,showpacs,twocolumn,floatfix,nofootinbib]{revtex4-1}
\usepackage[dvipdfmx]{graphicx}
\usepackage{amsmath,amssymb,siunitx}
\usepackage{color}

\begin{document}
\title{Properties of the remnant disk and the dynamical ejecta produced in low-mass black hole-neutron star mergers }
\author{
  Kota Hayashi$^{1}$,
  Kyohei Kawaguchi$^{2}$,
  Kenta Kiuchi$^{3,1}$,
  Koutarou Kyutoku$^{4,1,5}$,
  and
  Masaru Shibata$^{3,1}$
}
\affiliation{
  $^1$Center for Gravitational Physics, Yukawa Institute for Theoretical Physics,
  Kyoto University, Kyoto 606-8502, Japan\\
  $^2$Institute for Cosmic Ray Research,
  The University of Tokyo, 5-1-5 Kashiwanoha, Kashiwa, Chiba 277-8582, Japan\\
  $^3$Max Planck Institute for Gravitational Physics (Albert Einstein Institute),
  Am M{\"u}hlenberg 1, Postdam-Golm 14476, Germany\\
  $^4$Department of Physics, Kyoto University, Kyoto 606-8502, Japan\\
  $^5$Interdisciplinary Theoretical and Mathematical Sciences Program (iTHEMS), RIKEN, Wako, Saitama 351-0198, Japan
}
\date{\today}

\begin{abstract}
  We systematically perform numerical-relativity simulations for
  low-mass black hole-neutron star mergers for the models with seven mass ratios $Q=M_{\rm BH}/M_{\rm NS}$ ranging from 1.5 to 4.4,
  and three neutron-star equations of state,
  focusing on properties of matter remaining outside the black hole and
  ejected dynamically during the merger.
  We pay particular attention to the dependence on the mass ratio of the binaries.
  It is found that the rest mass remaining outside the apparent horizon after the merger depends only weakly on the mass ratio for the models with low mass ratios.
  It is also clarified that the rest mass of the ejecta has a peak at $Q \sim 3$,
  and decreases steeply as the mass ratio decreases for the low mass-ratio case. 
  We present a novel analysis method for the behavior of matter during the merger,
  focusing on the matter distribution in the phase space of specific energy and specific angular momentum.
  Then we model the matter distribution during and after the merger.
  Using the result of the analysis, we discuss the properties of the ejecta.
\end{abstract}

\maketitle

\section{Introduction} \label{sec:intro}
The era of the gravitational-wave astronomy was opened
by the first detection of a binary black hole merger GW150914 \cite{abbott2016feb}.
For the event GW170817 \cite{abbott2017oct1},
the electromagnetic counterpart signals were successfully detected \cite{abbott2017oct2, abbott2017oct3}.
It is widely believed that the source of GW170817 is the merger of binary neutron stars.
However, the fact that became clear about the source of GW170817 is limited as follows.
The observation of gravitational waves indicates that the component masses are $1.36$--$1.89 M_{\odot}$ and $1.00$--$1.36 M_{\odot}$ \cite{abbott2019jan}.
The observations of a short gamma-ray burst GRB 170817A and a kilonova/macronova AT 2017gfo
by electromagnetic waves indicate that matter is involved in
the merger process and at least one of the compact objects consisting of the binary is a neutron star.
On the basis of these observational facts, however, the possibility of the source being a black hole-neutron star merger is not completely excluded \cite{foucart2019may,hinderer2019sep}.
In addition, GW190425, an event detected in 2019, is analyzed to be the gravitational wave radiated from
the merger of binary compact objects, the component masses of which are
$1.61$--$2.52 M_{\odot}$ and $1.12$--$1.68 M_{\odot}$ \cite{abbott2020jan}.
It has not been determined whether the source is a black hole-neutron star merger or a binary neutron star merger for this event either \cite{kyutoku2020feb}.
Including these two events, there are increasing numbers of gravitational-wave event candidates
and some of them are thought to include neutron stars,
but the source objects of these events are not strongly constrained.

In order to determine the properties of the sources, it is crucial to understand the behavior of the system theoretically.
High-accuracy gravitational-wave templates are needed for the parameter estimation after the detection of gravitational waves,
and models for electromagnetic emissions are essential tools for analyzing electromagnetic counterpart signals.
The key quantities for determining electromagnetic emissions include
the remnant disk mass, the ejecta mass, and the ejecta velocity.
Here, the ejecta is the matter which becomes unbound from the system.
{\it R}-process nucleosynthesis in the neutron-rich matter such as the matter outflowed from the remnant accretion disk and the dynamically ejected matter
is expected to power an electromagnetic transient kilonova/macronova \cite{li1998nov, metzger2010jun}.
Also, the compact object surrounded by an accretion disk has been proposed
as a likely candidate for the central engine of short gamma-ray bursts (see Refs.~\cite{nakar2007apr, berger2014jun} for reviews).

A variety of numerical-relativity simulations have been performed for black hole-neutron star binaries
\cite{kyutoku2010aug, kyutoku2011sep, kyutoku2013aug, kyutoku2015aug, shibata2006dec, shibata2008apr,shibata2007may,etienne2008apr, duez2008nov, shibata2009feb, etienne2009feb, chawla2010sep, duez2010may, foucart2011jan, foucart2012feb, etienne2012mar, etienne2012oct, foucart2013apr, lovelace2013jun, deaton2013sep, foucart2014jul, paschalidis2015jun, kawaguchi2015jul, kiuchi2015sep, foucart2017jan, kyutoku2018jan, ruiz2018dec, foucart2019feb, foucart2019may, hinderer2019sep},
and quantitative dependence of the merger behavior on binary parameters has been extensively studied.
Some work took into account magnetic fields \cite{chawla2010sep, etienne2012mar, etienne2012oct, paschalidis2015jun, kiuchi2015sep, ruiz2018dec},
nuclear-theory-based equations of state (EOSs)
\cite{duez2010may, kyutoku2013aug, kyutoku2015aug, deaton2013sep, foucart2014jul, kawaguchi2015jul, kiuchi2015sep, foucart2017jan, kyutoku2018jan, foucart2019feb, foucart2019may, hinderer2019sep},
neutrino cooling
\cite{deaton2013sep, foucart2014jul, foucart2017jan, kyutoku2018jan, foucart2019may, hinderer2019sep},
neutrino heating
\cite{kyutoku2018jan},
and misalignment of the black-hole spin with respect to the orbital angular momentum
\cite{foucart2011jan, foucart2013apr, kawaguchi2015jul, foucart2017jan}.
However, previous studies on the black hole-neutron star mergers which took nuclear-theory-based EOSs into account
focused primarily on the system with a black-hole mass larger than $5 M_{\odot}$.
There are only a small number of studies carried out for low-mass black hole-neutron star mergers 
with black-hole mass of $2$--$3 M_{\odot}$ \cite{foucart2019feb}.
This is because the black-hole mass observed in our galaxy was in the range of  $5$--$20 M_{\odot}$ \cite{zel2012sep},
and a black hole with a mass lower than $\sim 3 M_{\odot}$ was not highly expected to exist.
However, recent electromagnetic observations are indicating the existence of
a compact object with mass $\sim 3.3 M_{\odot}$ \cite{thompson2019nov}.
Also, some gravitational-wave event candidates indicate the existence of a compact object with mass
in the range of the mass gap of $2$--$5 M_{\odot}$ \cite{abbott2020jun, gracedb}.
In this situation, extensive theoretical studies on low-mass black hole-neutron star mergers are urgently needed.

In the present study, we perform numerical relativity simulations of low-mass black hole-neutron star mergers.
The simulations were performed systematically in order to study the parameter dependence of the merger remnant of the system. 
Specifically, the simulations are performed for seven initial black-hole mass $M_{\rm BH}=2.0, 2.5, 3.0, 3.5, 4.0, 5.0,$ and $6.0 M_{\odot}$
and three neutron-star EOSs.
We pay particular attention to the rest mass remaining outside the apparent horizon after the merger, the ejecta mass, and the ejecta velocity.

This paper is organized as follows.
In Sec.~\ref{sec:methods}, we briefly summarize the method for the numerical simulation and the diagnostics.
In Sec.~\ref{sec:results}, we present the numerical results from the simulations
focusing on the dependence of the rest mass remaining outside the apparent horizon after the merger, the ejecta mass, and the ejecta velocity
on the mass ratio and the neutron-star EOS.
Finally, a conclusion of this work is presented in Sec.~\ref{sec:conclusion}. 
Throughout this paper, we adopt the geometrical units in which $G = c = 1$,
where $G$ and $c$ are the gravitational constant and the speed of light, respectively.
Our convention of notation is summarized in Table \ref{tab:notation}.
The compactness of the neutron star,
the total mass of the system at infinite separation,
and the mass ratio of the binary
are defined as
$\mathcal{C}:=M_{\rm NS}/R_{\rm NS}$, $m_0:=M_{\rm BH}+M_{\rm NS}$, and $Q:=M_{\rm BH}/M_{\rm NS}$,
respectively.
Latin and Greek indices denote
spatial and spacetime components, respectively.

\begin{table}[]
  \centering

  \caption{
    Our convention of notation for physically important quantities, geometric variables, and hydrodynamic variables.
  }
  \label{tab:notation}
  
  \begingroup
  \setlength{\tabcolsep}{2pt} 
  \renewcommand{\arraystretch}{1.2} 
  
  \begin{tabular}{c c}
    \hline
    Symbol & \\
    \hline
    \hline
    $M_{\rm BH}$ & Gravitational mass of the black hole in isolation \\
    $a_{\rm BH}$ & Kerr parameter of the black hole \\
    $\chi_{\rm BH}$ & Dimensionless spin parameter of the black hole \\
    $M_{\rm NS}$ & Gravitational mass of the neutron star in isolation \\
    $R_{\rm NS}$ & Circumferential radius of the neutron star in isolation \\
    $\mathcal{C}$ & Compactness parameter of the neutron star: $M_{\rm NS}/R_{\rm NS}$ \\
    $m_0$ & Total mass of the system at infinite separation \\
    $Q$ & Mass ratio of the binary:  $M_{\rm BH}/M_{\rm NS}$\\
    \hline
    $\gamma_{ij}$ &    Induced metric on a $t={\rm const.}$ hypersurface \\
    $\alpha$ &         Lapse function \\
    $\beta^i$ &        Shift vector \\
    $\gamma$ &         Determinant of $\gamma_{ij}$ \\ 
    \hline
    $\rho$ &           Baryon rest-mass density  \\
    $u^{\mu}$ &         Four velocity of the fluid \\
    $P$ &              Pressure \\
    $\varepsilon$ &       Specific internal energy \\
    $h$ &              Specific enthalpy: $1+\varepsilon+P/\rho$ \\
    $\rho_*$ &         Conserved baryon rest-mass density: $\rho \alpha \sqrt{\gamma} u^t$\\
    $\hat{e}$ &        Specific energy: $h \alpha u^t -P/\rho \alpha u^t $\\
    \hline
  \end{tabular}
  
  \endgroup

\end{table}

\section{Methods} \label{sec:methods}
In this section, we present methods for the numerical simulation. 
The details of the formulation, the gauge conditions, the numerical scheme, and the initial data computation are described in
Refs.~\cite{kyutoku2010aug,kyutoku2011sep,kawaguchi2015jul,kyutoku2015aug}.

\subsection{Dynamical simulation} \label{sec:dyn}
Numerical simulations are carried out by using the SACRA-MPI code \cite{kiuchi2017oct}.
This code employs an adaptive-mesh-refinement (AMR) method to save the computational cost \cite{yamamoto2008sep} and
MPI/OpenMP hybrid parallelization to speed up the computation \cite{kiuchi2017oct}. 
SACRA solves the Einstein equation 
in a moving puncture version \cite{campanelli2006mar,baker2006mar,marronetti2008mar} of the Baumgarte-Shapiro-Shibata-Nakamura (BSSN) formulation \cite{shibata1995nov,baumgarte1998dec},
incorporating a Z4c consrtraint-propagation prescription locally \cite{hilditch2013oct}.
Together with the Einstein equation, we solve pure hydrodynamics equations in this paper.
Magnetohydrodynamics or neutrino effects are not taken into account 
because we focus on the dynamics of the system up to $\sim \SI{15}{\ms}$ after the merger. 

In this work, we prepare ten refinement levels for the AMR computational domain.
Specifically, two sets of four finer domains comoving with either the black hole or the neutron star cover the region of their vicinity.
The other six coarser domains cover both the black hole and the neutron star by a wider domain
with their origins fixed at the center of the mass of the binary system.


\subsection{Zero-temperature EOS}
The temperature of the neutron star,
except for the newly born ones and the massive neutron star produced after the merger of binary neutron stars,
can be approximated as zero
because the cooling time scale of a neutron star is much shorter than the lifetime of typical compact binaries \cite{lattimer2004apr}.
With a zero-temperature EOS, thermodynamical quantities such as pressure $P$, specific internal energy $\varepsilon$, and
specific enthalpy $h$ are written as a function of rest-mass density $\rho$.
In this study we employ a piecewise polytropic EOS with two pieces as \cite{read2009jun, read2009jun2, lackey2012feb}
\begin{eqnarray} \label{}
  P_{\rm cold}(\rho)=K_{i} \rho^{\Gamma_{i}}
  &&  \quad \ (\rho_{i-1}<\rho<\rho_{i}, \ i=1,2) 
\end{eqnarray}
where $\rho_0=0$, $\rho_1$ is of order $\SI{e14}{\gram\per\cubic\cm}$ (see below), and $\rho_2=\infty$.  
Below $\rho_1$, we adopt
$K_1=3.5966 \times 10^{13}$ in cgs unit, $\Gamma_1=1.3569$,
and for $\rho \geq \rho_1$, we adopt $\Gamma_2=3.0$.
The remaining free parameters are $K_2$ and $\rho_1$.
These are determined by choosing pressure $P_{\rm fid}$ at certain fiducial density $\rho_{\rm fid}=\SI{e14.7}{\gram\per\cubic\cm}$
as $K_2=P_{\rm fid}/{\rho_{\rm fid}}^{\Gamma_2}$ 
and by requiring the continuity of the pressure as $\rho_1=(K_2/K_1)^{1/(\Gamma_1-\Gamma_2)}$.
We choose three EOSs shown in Table \ref{tab:ns-eos}
in order to investigate the EOS dependence of the merger outcome of the binary systems.
These selected EOSs satisfy the constraints $\Lambda_{1.4}\lesssim 800$ imposed by the observation of GW170817 \cite{abbott2017oct1}.
Here, $\Lambda_{1.4}$ is the dimensionless tidal deformability for an isolated neutron star with the gravitational mass $1.4 M_{\odot}$.
Also with these EOSs, the maximum mass for the spherical neutron stars, $M_{\rm max}$ is higher than $2.1 M_{\odot}$, 
which is consistent with the latest discoveries of a high-mass neutron star \cite{cromartie2019sep}.
In the numerical simulation, we add the thermal part of the EOS to the zero-temperature part described above. 
Our implementation for this is the same as that in our previous work (see, e.g., Ref.~\cite{kawaguchi2015jul}). 

\begin{table}[]
  \centering

  \caption{
    The EOS parameter adopted in this study.
    $M_{\rm max}$ is the maximum gravitational mass of the spherical neutron star for each EOS.
    $R_{1.35}$ and $\mathcal{C}_{1.35}$ are the radius and the compactness
    for a neutron star with the gravitational mass $1.35 M_{\odot}$, respectively.
  }
  \label{tab:ns-eos}
  
  \begingroup
  \setlength{\tabcolsep}{4pt} 
  \renewcommand{\arraystretch}{1.2} 
  
  \begin{tabular}{c c c c c}
    \hline
    \hline
    EOS   & $\log_{10}P_{\rm fid} [{\rm dyne/cm^{2}}]$ & $M_{\rm max} [M_{\odot}]$  & $R_{1.35} [{\rm km}]$ & $\mathcal{C}_{1.35}$ \\
    \hline
    1.25H & 34.6                 & 2.383         & 13.0                 & 0.154               \\ 
    H     & 34.5                 & 2.249         & 12.3                 & 0.162               \\ 
    HB    & 34.4                 & 2.122         & 11.6                 & 0.172               \\
    \hline
    \hline
  \end{tabular}
  
  \endgroup

\end{table}

\subsection{Model}
The neutron-star mass is set to be $1.35 M_{\odot}$ with the vanishing black-hole spin for all the models.
In order to investigate the mass-ratio dependence of the system evolution,
we choose seven mass ratios $Q\approx 1.5, 1.9, 2.2, 2.6, 3.0, 3.7,$ and $4.4$,
that is, we choose the black-hole mass $M_{\rm BH,0}= 2.0, 2.5, 3.0, 3.5, 4.0, 5.0,$ and $6.0 M_{\odot}$.
As mentioned above, we also choose three EOSs 1.25H, H, and HB.
The 19 physical models used in this study are listed in Table \ref{tab:models}.
Strictly speaking, the neutron star EOS with $M_{\rm max}>2.1M_{\odot}$ and $M_{\rm BH}=2M_{\odot}$ are astrophysically inconsistent. 
However, we adopt this value because the main focus of this paper is to understand the physics of the low-mass black hole-neutron star merger process.

Initial data for our numerical simulations are obtained by
computing a quasi-equilibrium state of an orbiting black hole-neutron star binaries following Ref.~\cite{kyutoku2009jun}.
Here, we assume that the neutron star is irrotational \cite{bildsten1992nov,kochanek1992oct} and
is modeled by zero-temperature EOSs (see Ref.~\cite{lattimer2004apr} for reviews).

In order to confirm the reliability of the numerical results,
we perform three simulations for every model listed above with different grid resolutions N70, N90, and N110.
They resolve the neutron-star radius by about 50, 65, and 80 grid points on the finest AMR domain, respectively.
The grid spacing for the highest-resolution model N110 is $\Delta x \approx 110$--$\SI{130}{m}$ on the finest AMR domain.
Unless otherwise stated, the numerical results from the simulations of N110 are presented for each model. 

\begin{table*}[h]
  \centering
  
  \caption{
    Key parameters and quantities for the initial conditions adopted in our numerical simulations.
    The adopted EOS, the compactness of the neutron star $\mathcal{C}$, and the initial black-hole mass $M_{\rm BH,0}$ are shown.
    Note that $M_{\rm NS}=1.35 M_{\odot}$ and the initial black-hole spin is zero.
    $m_0\Omega_0$, $M_{\rm ADM,0}$ , and $J_{\rm ADM,0}$
    are the initial dimensionless orbital angular velocity,
    Arnowitt-Deser-Misner (ADM) mass, and ADM angular momentum of the system, respectively.
    $\Delta x$ is the grid spacing for the highest resolution model N110 and $L$ is the size of the computational domain.
  }
  \label{tab:models}
  
  \begingroup
  \setlength{\tabcolsep}{8pt} 
  \renewcommand{\arraystretch}{1.2} 
  
  \begin{tabular}{c ccc ccc cc}
    \hline
    \hline
    Model & EOS   & $\mathcal{C}$ & $M_{\rm BH,0} [M_{\odot}]$ & $m_0 \Omega_{\rm 0}$ & $M_{\rm ADM,0} [M_{\odot}]$ & $J_{\rm ADM,0} [M_{\odot}^2]$ & $\Delta x [{\rm m}]$ & $L [{\rm km}]$\\
    \hline
    125H\_Q15 & 1.25H & 0.154 & 2.0  & 0.024 & 3.33 & 10.63 & 127 & 7206\\
    125H\_Q19 & 1.25H & 0.154 & 2.5  & 0.024 & 3.83 & 13.27 & 124 & 7006\\
    125H\_Q22 & 1.25H & 0.154 & 3.0  & 0.024 & 4.33 & 15.90 & 124 & 7006\\
    125H\_Q26 & 1.25H & 0.154 & 3.5  & 0.024 & 4.83 & 18.53 & 124 & 7006\\
    125H\_Q30 & 1.25H & 0.154 & 4.0  & 0.024 & 5.34 & 21.15 & 124 & 7006\\
    125H\_Q37 & 1.25H & 0.154 & 5.0  & 0.026 & 6.34 & 25.88 & 122 & 6906\\
    125H\_Q44 & 1.25H & 0.154 & 6.0  & 0.026 & 7.35 & 30.97 & 122 & 6906\\   \hline              
    H\_Q15    & H     & 0.162 & 2.0  & 0.024 & 3.33 & 10.63 & 117 & 6606\\
    H\_Q19    & H     & 0.162 & 2.5  & 0.024 & 3.83 & 13.27 & 117 & 6606\\
    H\_Q22    & H     & 0.162 & 3.0  & 0.024 & 4.33 & 15.91 & 117 & 6606\\
    H\_Q26    & H     & 0.162 & 3.5  & 0.024 & 4.83 & 18.53 & 117 & 6606\\
    H\_Q30    & H     & 0.162 & 4.0  & 0.024 & 5.34 & 21.15 & 117 & 6606\\
    H\_Q37    & H     & 0.162 & 5.0  & 0.026 & 6.34 & 25.88 & 113 & 6406\\
    H\_Q44    & H     & 0.162 & 6.0  & 0.026 & 7.35 & 30.97 & 122 & 6406\\   \hline             
    HB\_Q15   & HB    & 0.172 & 2.0  & 0.024 & 3.33 & 10.63 & 109 & 6139\\
    HB\_Q19   & HB    & 0.172 & 2.5  & 0.024 & 3.83 & 13.27 & 109 & 6139\\
    HB\_Q22   & HB    & 0.172 & 3.0  & 0.024 & 4.33 & 15.91 & 109 & 6139\\
    HB\_Q26   & HB    & 0.172 & 3.5  & 0.024 & 4.83 & 18.53 & 109 & 6139\\
    HB\_Q30   & HB    & 0.172 & 4.0  & 0.024 & 5.34 & 21.15 & 109 & 6139\\
    \hline
    \hline
  \end{tabular}

  \endgroup

\end{table*}

\subsection{Diagnostics} \label{sec:diag}
\subsubsection{Remnant disk and ejecta}
The fate of the neutron-star matter after merger is divided into three types.
The matter that falls immediately into the black hole,
the matter that forms an accretion disk,
and the matter that becomes unbound from the system, i.e., ejecta.
Here, we describe our method to evaluate the properties of the disk and the ejecta,
which are the key quantities for the electromagnetic emissions from black hole-neutron star mergers.
At each time slice, the rest mass outside the apparent horizon is evaluated by the integral
\begin{eqnarray}
  M_{> {\rm AH}}:=\int_{r>r_{\rm AH}}\rho_*d^3x ,
\end{eqnarray}
where $r_{\rm AH}=r_{\rm AH}(\theta, \varphi)$ is the coordinate radius of the apparent horizon
with $\theta$ and $\varphi$ being the polar angles defined in a black-hole centered frame.
Here, we define the time at the onset of merger $t_{\rm merger}$ as
the time at which $10^{-2} M_{\odot}$ of neutron-star matter falls into the apparent horizon.
$M_{> {\rm AH}}$ is evaluated at $\SI{12}{\ms}$ after the onset of merger.

The ejecta, which is the matter unbound from the system, is defined as the matter with $-u_{t}>1$.\footnote{
If we consider the thermal effect, the criterion for the unbound matter should be expressed as $-hu_{t}>1$.
For the dynamical ejecta produced in the merger of black hole-neutron star binaries,
matter is ejected mainly due to the tidal force and it does not experience the shock heating.
Therefore, numerical results in this paper depend only weakly on the choice of the criterion.}
The mass of the ejecta is defined by integrating the conserved rest-mass density of the matter with $-u_{t}>1$ as
\begin{eqnarray}
  M_{\rm eje}:=\int_{-u_{t}>1,r>r_{\rm AH}}\rho_*d^3x.
\end{eqnarray}
The average velocity of the ejecta is defined by considering the kinetic energy of the ejecta.
First, a sum of the rest-mass, internal and kinetic energies of the ejecta is defined by
\begin{eqnarray}
  E_{\rm eje}:=\int_{-u_{t}>1,r>r_{\rm AH}}\rho_*\hat{e}d^3x,
\end{eqnarray}
where $\hat{e}$ is defined in Table~\ref{tab:notation}.
Next, the internal energy of the ejecta is defined by
\begin{eqnarray}
  U_{\rm eje}:=\int_{-u_{t}>1,r>r_{\rm AH}}\rho_* \varepsilon d^3x.
\end{eqnarray}
Then, 
the kinetic energy of the ejecta is defined by
subtracting the rest mass energy and internal energy from $E_{\rm eje}$ as
\begin{eqnarray}
  T_{\rm eje}:=E_{\rm eje}-U_{\rm eje}-M_{\rm eje}.
\end{eqnarray}
Subsequently, by assuming the Newtonian dynamics,
average velocity of the ejecta may be evaluated from
its kinetic energy and the mass as
\begin{eqnarray}
  v_{\rm eje}:=\sqrt{ \frac{2T_{\rm eje}}{M_{\rm eje}} }.
\end{eqnarray}
However, the values of $E_{\rm eje}$ and $U_{\rm eje}$ are evaluated for a computational domain of radius $< 6000$--$\SI{7000}{km}$. 
Then, in $E_{\rm eje}$ (and thus $T_{\rm eje}$), the influence on the gravitational potential energy remains. 
We must subtract the effect of this gravitational potential energy.
Assuming the Newtonian gravity,
we thus approximately estimate the extrapolated velocity as
\begin{eqnarray}
  v_{\rm eje,extrap}:=\sqrt{ {v_{\rm eje}}^2-2\frac{m_0}{v_{\rm eje}(t-t_{\rm merger})} } \label{eq:veje_extrap},
\end{eqnarray}
where $v_{\rm eje}$ is evaluated at $t$.
$M_{\rm eje}$ and $v_{\rm eje}$ are also evaluated at $\SI{12}{\ms}$ after the onset of merger.


\subsubsection{Black hole}
Parameters of black holes are estimated by integrals on apparent horizons.
By assuming that the spacetime is approximately stationary and the effect of matter would be negligible near the black hole,
the equatorial circumferential radius $C_{\rm e}$ and the area $A_{\rm AH}$ of the apparent horizon approximately satisfy \cite{shibata2009feb}
\begin{eqnarray}
  C_{\rm e} &=& 4\pi M_{\rm BH} , \\
  A_{\rm AH} &=& 8 \pi M_{\rm BH} \left(M_{\rm BH}+\sqrt{M_{\rm BH}^2-a_{\rm BH}^2} \right) .
\end{eqnarray}
Therefore, the black-hole mass $M_{\rm BH}$ and the dimensionless spin parameter $\chi_{\rm BH}$
are approximately evaluated as
\begin{eqnarray}
  M_{\rm BH}  &=& \frac{C_{\rm e}}{4\pi} , \\
  \chi_{\rm BH} &=& \frac{1}{M_{\rm BH}} \sqrt{M_{\rm BH}^2 - \left( \frac{A_{\rm AH}}{8 \pi M_{\rm BH}} - M_{\rm BH} \right)^2 } .
\end{eqnarray}
Comparisons among different estimates of the spin parameter suggest that the systematic error associated with
this method is smaller than $0.01$ \cite{kyutoku2010aug,kyutoku2011sep,shibata2009feb}, and we confirmed that this also holds for simulations presented in this work.
As in the case of the remnant disk and ejecta, the mass and dimensionless spin of the black hole are
estimated at \SI{12}{\ms} after the onset of merger.

\subsubsection{Orbital angular velocity}
Here, we summarize the method for computing the orbital angular velocity from gravitational waves.
We extract a Weyl scalar $\Psi_4$ at the coordinate radius of $D = 400 M_{\odot}$ from the coordinate origin
by projecting onto spin-weighted spherical harmonics,
and extrapolate them to null infinity by a method based on the black-hole perturbation theory \cite{lousto2010nov}.
The gravitational waveforms $h_{\rm gw}$ are obtained from time integration of $l=2, m=\pm 2$ modes of $\Psi_4$ \cite{reisswig2011sep}.
The angular velocity of gravitational waves is derived by
\begin{eqnarray}
  \Omega_{\rm gw} = -\frac{1}{|h_{\rm gw}|^2} {\rm Im} \left[ h_{\rm gw}^{*} \dot{h}_{\rm gw}\right],
\end{eqnarray}
where $h_{\rm gw}^{*}$ and $\dot{h}_{\rm gw}$ denotes the complex conjugate and the time derivative of $h_{\rm gw}$, respectively.
The orbital angular velocity of the binary $\Omega$ is estimated as
$\Omega=\Omega_{\rm gw}/2$ in a gauge-invariant manner.
The retarded time is approximately defined by
\begin{eqnarray}
  t_{\rm ret} := t-D-2m_0 \ln( D/m_0 ).
\end{eqnarray}
Then we define the orbital angular velocity at the onset of merger as $\Omega_{\rm merger}:=\Omega(t_{\rm ret}=t_{\rm merger})$. 


\section{Results} \label{sec:results}
In this section, we present the results obtained from our numerical simulations.
Key characteristic quantities are shown in Table \ref{tab:rem_eje}.

\begin{table}[h]
  \centering

  \caption{
    Characteristic physical quantities of the remnant disk and the ejecta 
    measured at \SI{12}{\ms} after the onset of merger for models with the highest resolution N110.
    $M_{> {\rm AH}}$ is the rest mass outside the apparent horizon.
    $M_{\rm eje}$ is the rest mass of the ejecta.
    $v_{\rm eje,extrap}$ is the average velocity of the ejecta extrapolated to $r \to \infty$.
    For low ejecta mass $<10^{-5}M_{\odot}$, we do not estimate the ejecta velocity. 
  }
  \label{tab:rem_eje}
  
  \begingroup
  \setlength{\tabcolsep}{8pt} 
  \renewcommand{\arraystretch}{1.2} 
  
  \begin{tabular}{c ccc}
    \hline
    \hline
    Model & $M_{> {\rm AH}} [M_{\odot}]$ & $M_{\rm eje} [M_{\odot}]$ & $v_{\rm eje,extrap} [c]$   \\
    \hline
    125H\_Q15 & 1.0$\times 10^{-1}$ & 3.7$\times 10^{-4}$ & 0.10 \\
    125H\_Q19 & 1.0$\times 10^{-1}$ & 1.5$\times 10^{-3}$ & 0.12 \\
    125H\_Q22 & 1.0$\times 10^{-1}$ & 2.9$\times 10^{-3}$ & 0.14 \\
    125H\_Q26 & 9.1$\times 10^{-2}$ & 2.9$\times 10^{-3}$ & 0.14 \\
    125H\_Q30 & 6.7$\times 10^{-2}$ & 3.8$\times 10^{-3}$ & 0.16 \\
    125H\_Q37 & 1.4$\times 10^{-2}$ & 3.5$\times 10^{-3}$ & 0.18 \\
    125H\_Q44 & 3.5$\times 10^{-4}$ & 6  $\times 10^{-5}$ & 0.14 \\
    \hline                                           
    H\_Q15    & 7.0$\times 10^{-2}$ & 3  $\times 10^{-5}$ & 0.11 \\
    H\_Q19    & 7.3$\times 10^{-2}$ & 3.9$\times 10^{-4}$ & 0.12 \\
    H\_Q22    & 6.6$\times 10^{-2}$ & 8.8$\times 10^{-4}$ & 0.14 \\
    H\_Q26    & 4.4$\times 10^{-2}$ & 9.0$\times 10^{-4}$ & 0.13 \\
    H\_Q30    & 2.3$\times 10^{-2}$ & 2.0$\times 10^{-3}$ & 0.15 \\
    H\_Q37    & 1.3$\times 10^{-3}$ & 2.3$\times 10^{-4}$ & 0.15 \\
    H\_Q44    & $< 10^{-5}$         & $< 10^{-5}$         & -    \\
    \hline                                           
    HB\_Q15   & 3.7$\times 10^{-2}$ & $< 10^{-5}$         & -    \\
    HB\_Q19   & 3.9$\times 10^{-2}$ & 6  $\times 10^{-5}$ & 0.16 \\
    HB\_Q22   & 2.5$\times 10^{-2}$ & 9  $\times 10^{-5}$ & 0.11 \\
    HB\_Q26   & 1.6$\times 10^{-2}$ & 8.4$\times 10^{-4}$ & 0.14 \\
    HB\_Q30   & 5.0$\times 10^{-3}$ & 5.2$\times 10^{-4}$ & 0.14 \\
    \hline
    \hline
  \end{tabular}
  
  \endgroup

\end{table}

\subsection{Overview of the merger process} \label{sec:overview}
Here, we overview merger processes
(see Ref.~\cite{shibata2011aug} for reviews).
Gravitational radiation dissipates energy and angular momentum from black hole-neutron star binaries,
and the orbital separation decreases leading to merger.
The fate of the system is divided broadly into two cases.
One is that the neutron star is disrupted by the tidal force of the black hole.
If this is the case, the remnant black hole is expected to be surrounded by an accretion disk,
and a portion of the disrupted matter is ejected dynamically.
This is the case which we are interested in, and most of our models result in this type.
The other is that the neutron star is not tidally disrupted, and is simply swallowed by the black hole.
We do not pay particular attention to such models,
but this is the case for models with mass ratios $Q \gtrsim 4$--$5$.
The condition which discriminates these two cases is approximately obtained by comparing the orbital separation at which tidal disruption occurs
and the radius of the innermost stable circular orbit of the system.

\begin{figure*}[]
  \begin{tabular}{cc}
    
    \begin{minipage}[t]{0.5\hsize}
      \begin{center}
        \includegraphics[scale=0.45]{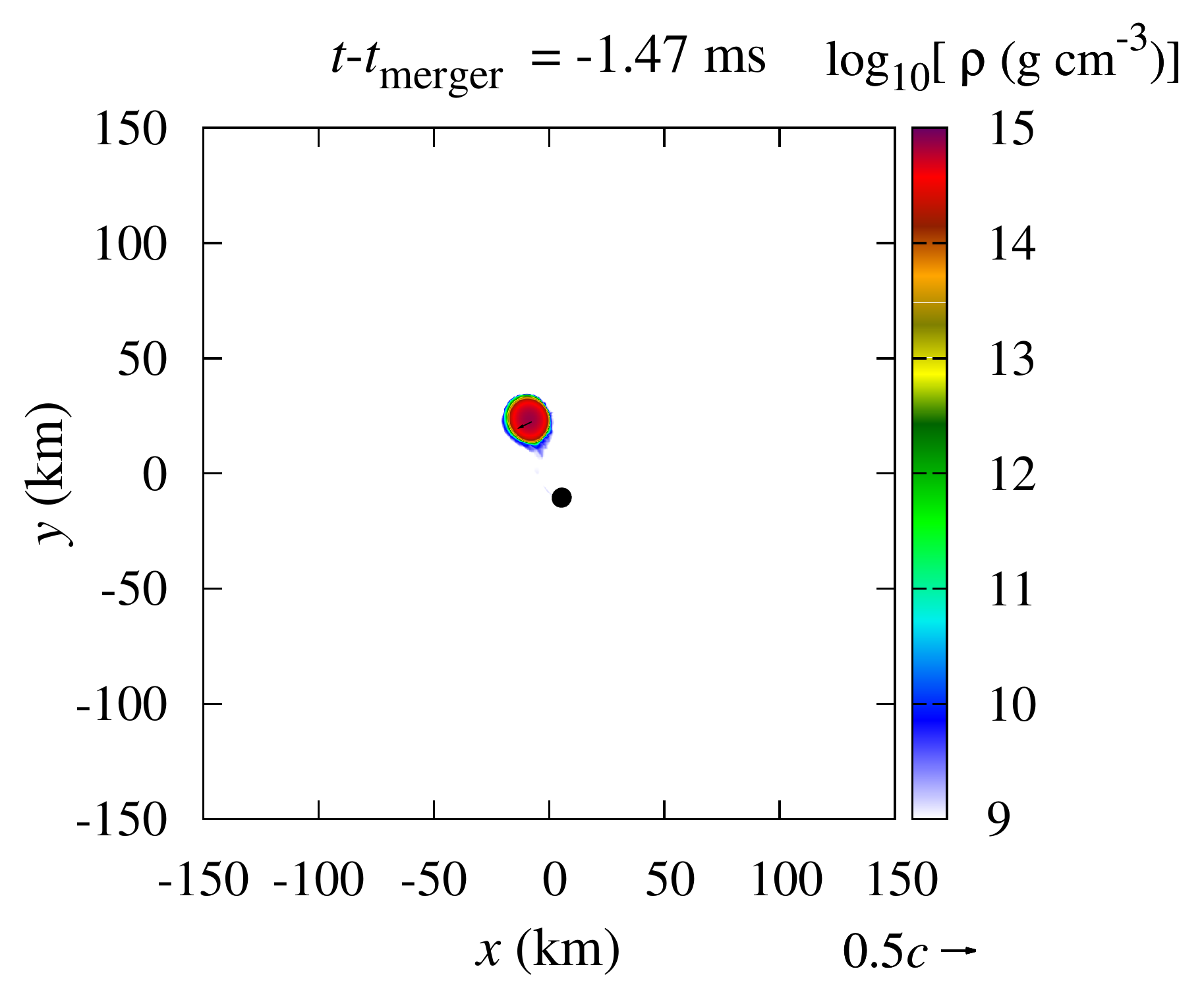}
      \end{center}
    \end{minipage}

    \begin{minipage}[t]{0.5\hsize}
      \begin{center}
        \includegraphics[scale=0.45]{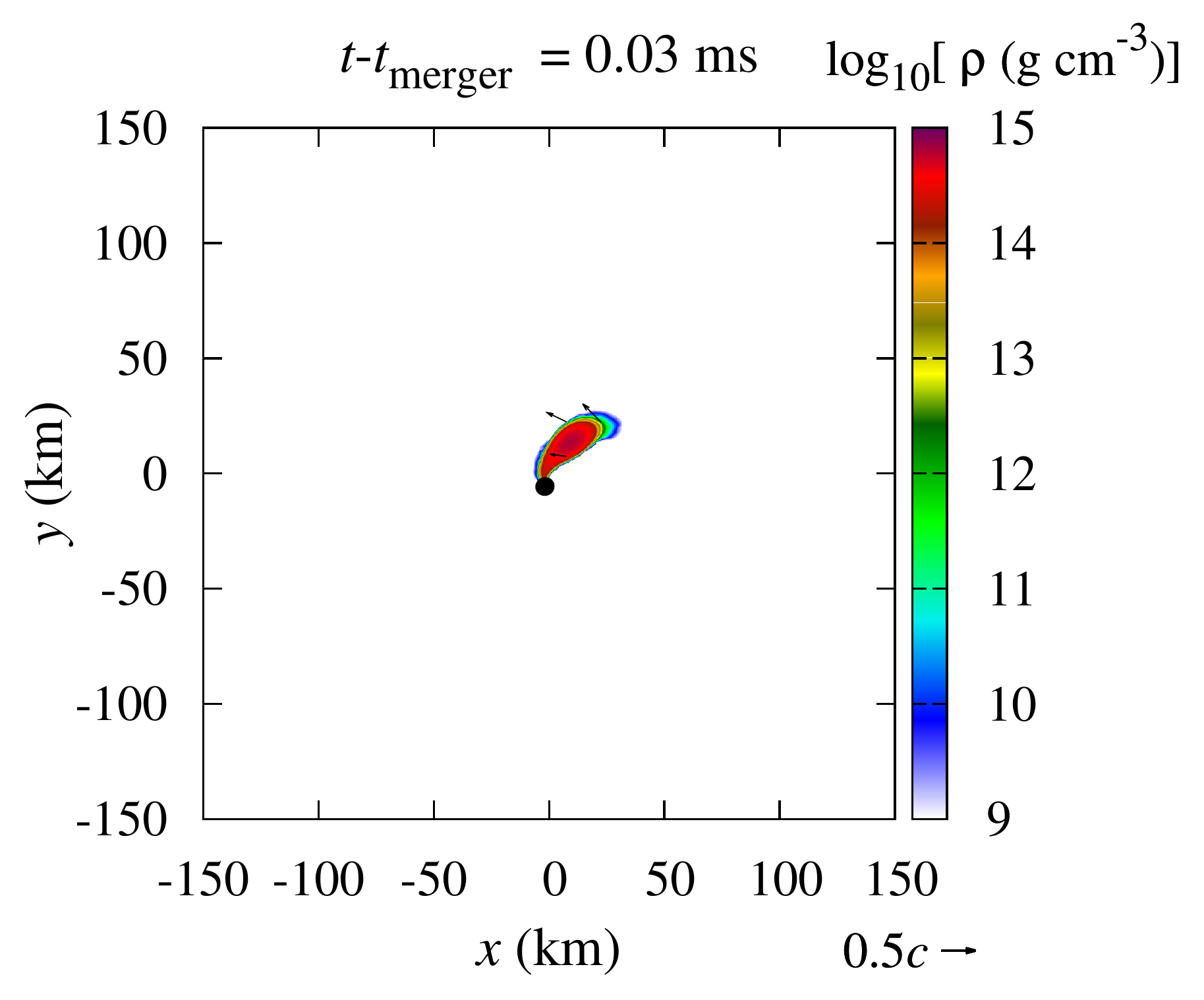}
      \end{center}
    \end{minipage}

  \end{tabular}


  \begin{tabular}{cc}
    
    \begin{minipage}[t]{0.5\hsize}
      \begin{center}
        \includegraphics[scale=0.45]{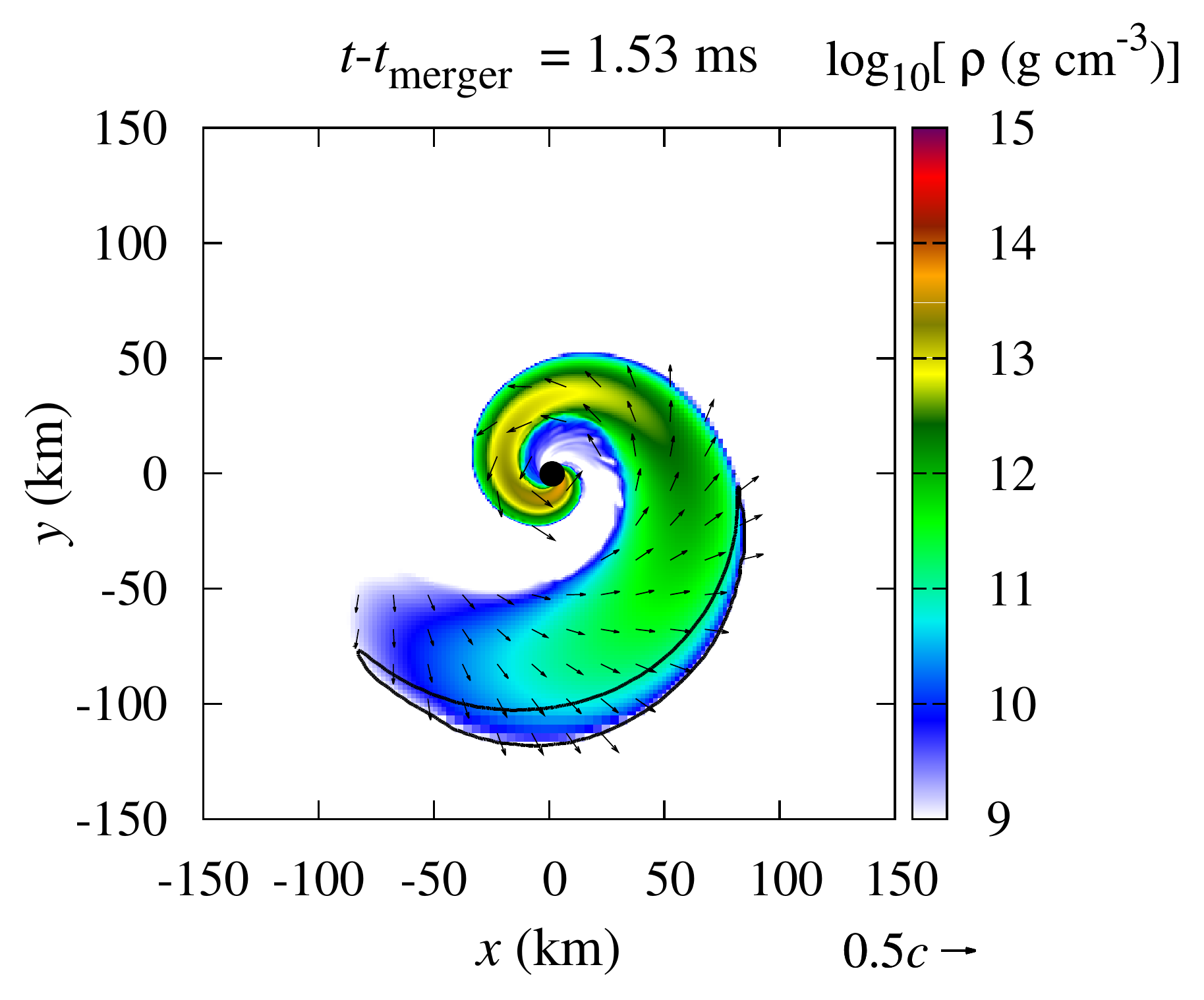}
      \end{center}
    \end{minipage}

    \begin{minipage}[t]{0.5\hsize}
      \begin{center}
        \includegraphics[scale=0.45]{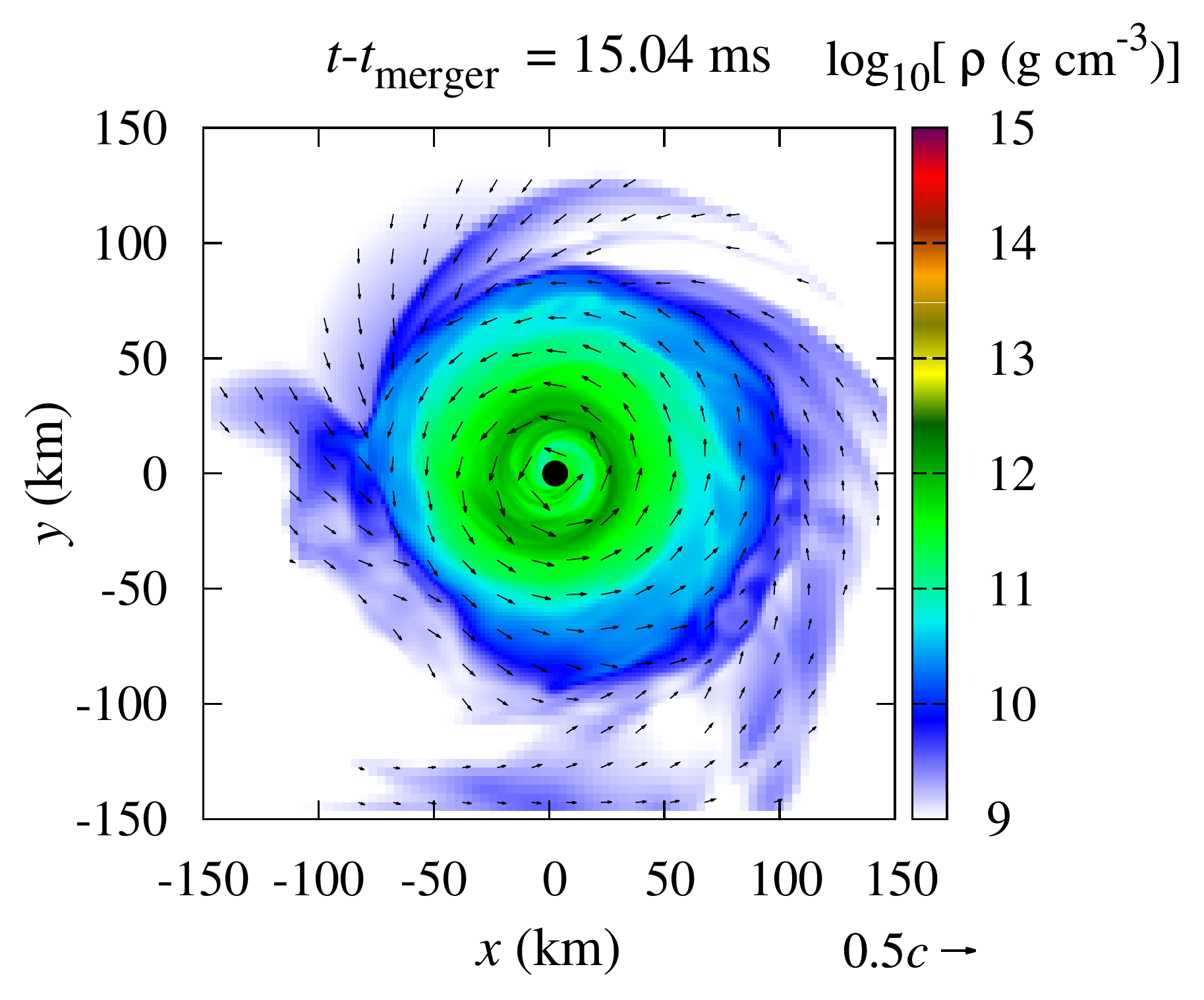}
      \end{center}
    \end{minipage}

  \end{tabular}

  \caption{
    Time evolution of the rest-mass density profile
    at $t-t_{\rm merger} \approx -1.5$ (top-left), $0.0$ (top-right), $1.5$ (bottom-left), and $\SI{15.0}{\ms}$ (bottom-right) for the model H\_Q22.
    The black filled circles indicate the interior of the apparent horizons.
    Unbound components that satisfy $-u_t \geq 1$ are enclosed by the black curves (see the bottom-left panel).
    The black arrows show the three-velocity, $v^i:=u^i/u^t$.
  }
   \label{fig:H_Q22_xy}

\end{figure*}

Figure \ref{fig:H_Q22_xy} plots the snapshots of the rest-mass density profiles,
unbound components, and the location of the apparent horizon on the equatorial plane for the model H\_Q22.
At $t-t_{\rm merger} \approx \SI{-1.5}{\ms}$,
the binary is in the inspiral stage, and the neutron star is not deformed appreciably. 
At $t-t_{\rm merger} \approx \SI{0.0}{\ms}$,  the binary is in the merger stage, 
and the neutron star is highly deformed by the tidal force of the black hole.
At $t-t_{\rm merger} \approx \SI{1.5}{\ms}$, we observe a one-armed spiral structure of the matter as a result of tidal disruption.
A large portion of the arm is kept bound to the remnant black hole,
and hence, they will experience fallback and result in the formation of an accretion disk around the remnant black hole.
On the other hand, a small portion of the arm at the front side acquires specific energy that satisfies $-u_t \geq 1$ and becomes ejecta.
At $t-t_{\rm merger} \approx \SI{15.0}{\ms}$, the accretion disk with the maximum density of $\sim \SI{e12}{\gram\per\cubic\cm}$
is formed around the remnant black hole.
A large portion of the disrupted matter is in a circular motion and the system relaxes to a quasi-steady state.

\subsection{Rest mass remaining outside the apparent horizon after the merger} \label{sec:rem_mass}
\begin{figure}[]
  \begin{tabular}{c}

    \begin{minipage}{1.0\hsize}
      \begin{center}
        \includegraphics[scale=0.35]{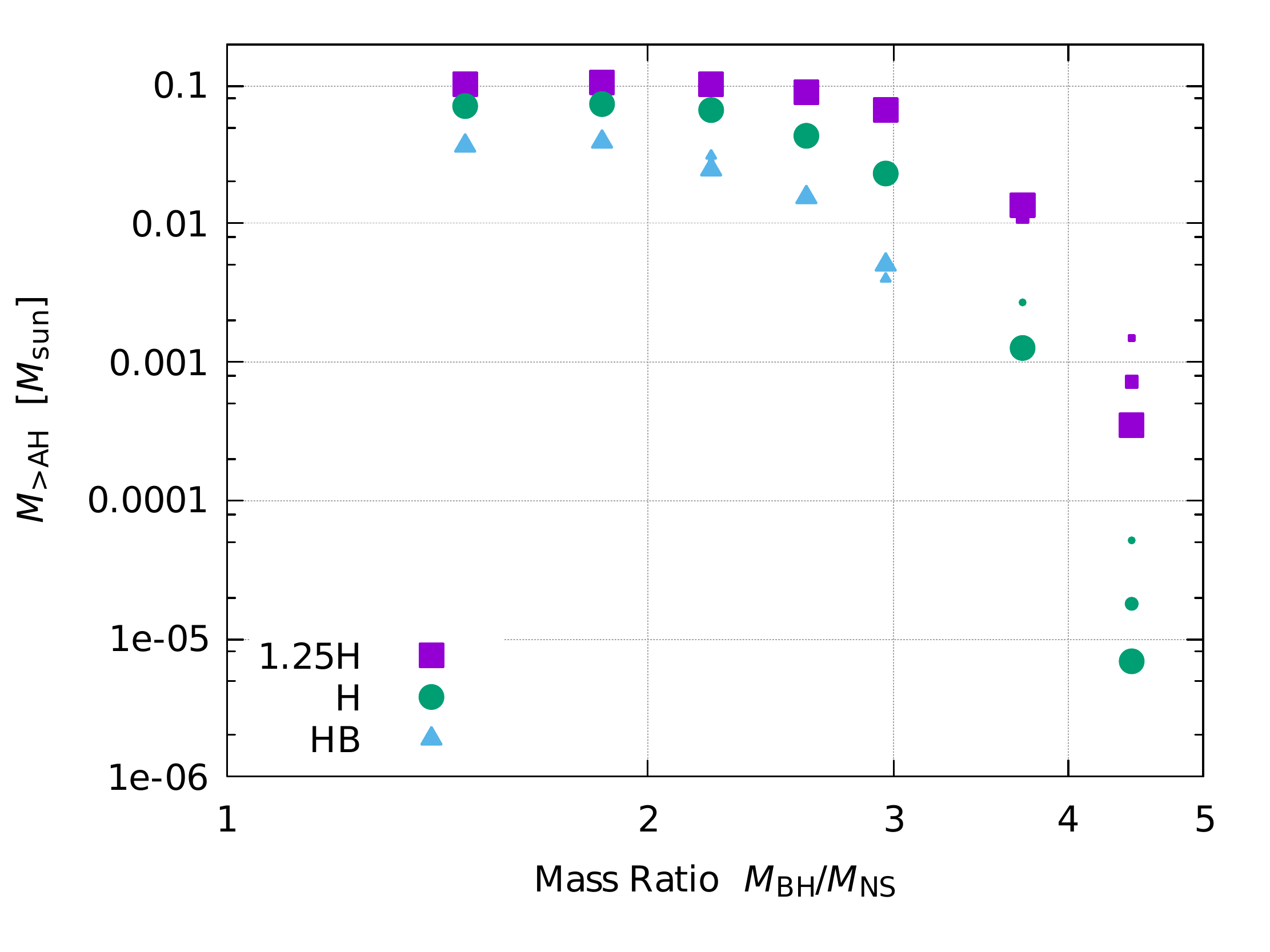}
        \caption{
          Rest mass remaining outside the apparent horizon after the merger $M_{> {\rm AH}}$ as a function of the mass ratio $Q$ for three EOSs.
          Data are evaluated at $\SI{12}{\ms}$ after the onset of merger.
          Data points with different sizes represent results obtained with different grid resolutions.
	  Specifically, large, medium, and small points show the results of N110, N90, and N70, respectively.
        }
        \label{fig:rem_er}
      \end{center}
    \end{minipage}

  \end{tabular}
\end{figure}

Figure \ref{fig:rem_er} shows the dependence of the rest mass remaining outside the apparent horizon after the merger $M_{> {\rm AH}}$ on the binary parameters.
First, we pay attention to the dependence of $M_{> {\rm AH}}$ on the EOS.
As we find from Fig.~\ref{fig:rem_er} and Table \ref{tab:rem_eje}, 
the rest mass remaining outside the apparent horizon after the merger increases as the compactness of the neutron star decreases.
This dependence is already shown by previous simulations \cite{kyutoku2010aug}.
This dependence reflects the fact that the neutron star with a larger radius is tidally disrupted at a more distant orbit.

Next, we pay attention to the dependence of $M_{> {\rm AH}}$ on the mass ratio of binary.
Figure \ref{fig:rem_er} shows that for high mass ratios ($Q \gtrsim 3$),
$M_{> {\rm AH}}$ increases as the mass ratio decreases.
This dependence was also already shown in previous simulations for a higher mass-ratio regime \cite{shibata2009feb}.
However, for a low mass ratio ($Q \lesssim 3$), 
the dependence of $M_{> {\rm AH}}$ on the mass ratio disappears.
This behavior was not expected from the mass-ratio dependence for the higher mass-ratio regime. 
It was already pointed out by Foucart et al. \cite{foucart2019may} that for a low-mass-ratio regime
$M_{> {\rm AH}}$ tends to be smaller than previously predicted in Ref.~\cite{foucart2012dec}.
Our work systematically shows that $M_{> {\rm AH}}$ becomes approximately constant irrespective of the mass ratio.

In Appendix \ref{app:rem_hQ_s}, we also reanalyze the results of our previous numerical simulations for models with higher mass ratios and spinning black holes \cite{kyutoku2011sep,kyutoku2015aug}. 
It is shown that the tendency similar to that shown in Fig.~\ref{fig:rem_er} is found for these previous results.

%
%

\subsection{Ejecta properties} \label{sec:eje_mass}
\begin{figure}[]
  \begin{tabular}{c}

    \begin{minipage}{1.0\hsize}
      \begin{center}
        \includegraphics[scale=0.35]{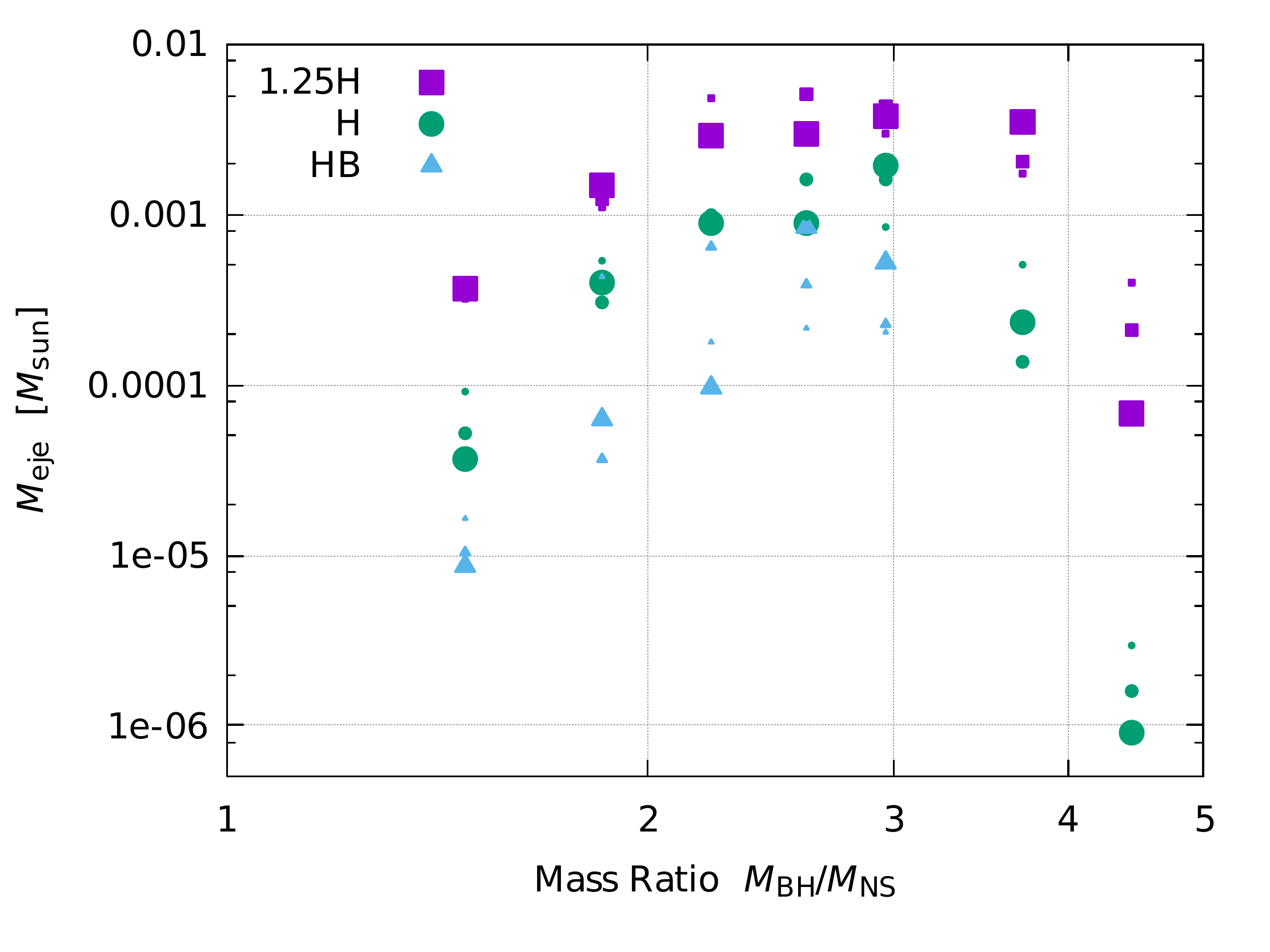}
        \caption{
          The same as Fig.~\ref{fig:rem_er} but for the rest mass of unbound material $M_{\rm eje}$ as a function of the mas ratio $Q$.
        }
        \label{fig:eje_er}
      \end{center}
    \end{minipage}

  \end{tabular}
\end{figure}

One of the important quantities that characterize ejecta is its mass. 
Figure \ref{fig:eje_er} shows the dependence of the ejecta mass $M_{\rm eje}$ on the mass ratio and neutron-star EOS.
It is found that as in the case of the rest mass remaining outside the apparent horizon after the merger, 
$M_{\rm eje}$ increases as the compactness of the neutron star decreases.
This dependence is consistent with the results in the previous work for higher mass ratios \cite{kyutoku2015aug}.  

For the models with the mass ratio higher than $\sim 3$,
$M_{\rm eje}$ decreases as the mass ratio increases.
This dependence is in agreement with the rapid drop of the rest mass remaining outside the apparent horizon after the merger at the higher mass-ratio regime.
This is also consistent with the results in the previous work \cite{kyutoku2015aug}.  
However, for the models with the mass ratio lower than $\sim 3$, the situation changes.
$M_{\rm eje}$ decreases as the mass ratio decreases in the parameter regime of our simulation.
This behavior is consistent with the numerical results in Ref.~\cite{foucart2019may},
which show that there is little unbound matter produced by a merger in the near-equal-mass regime.
Overall, we systematically show that $M_{\rm eje}$ exhibits a peak at the mass ratio $\sim 3$.
However, it is not evident that the peak always exists for the cases which are not explored in this work.
The peak mass ratio could depend on the neutron-star EOS and the black-hole spin.
Further study is needed to reveal whether the peak always exists or not.



\begin{figure}[]
  \begin{tabular}{c}
%
%

    \begin{minipage}{1.0\hsize}
      \begin{center}
        \includegraphics[scale=0.35]{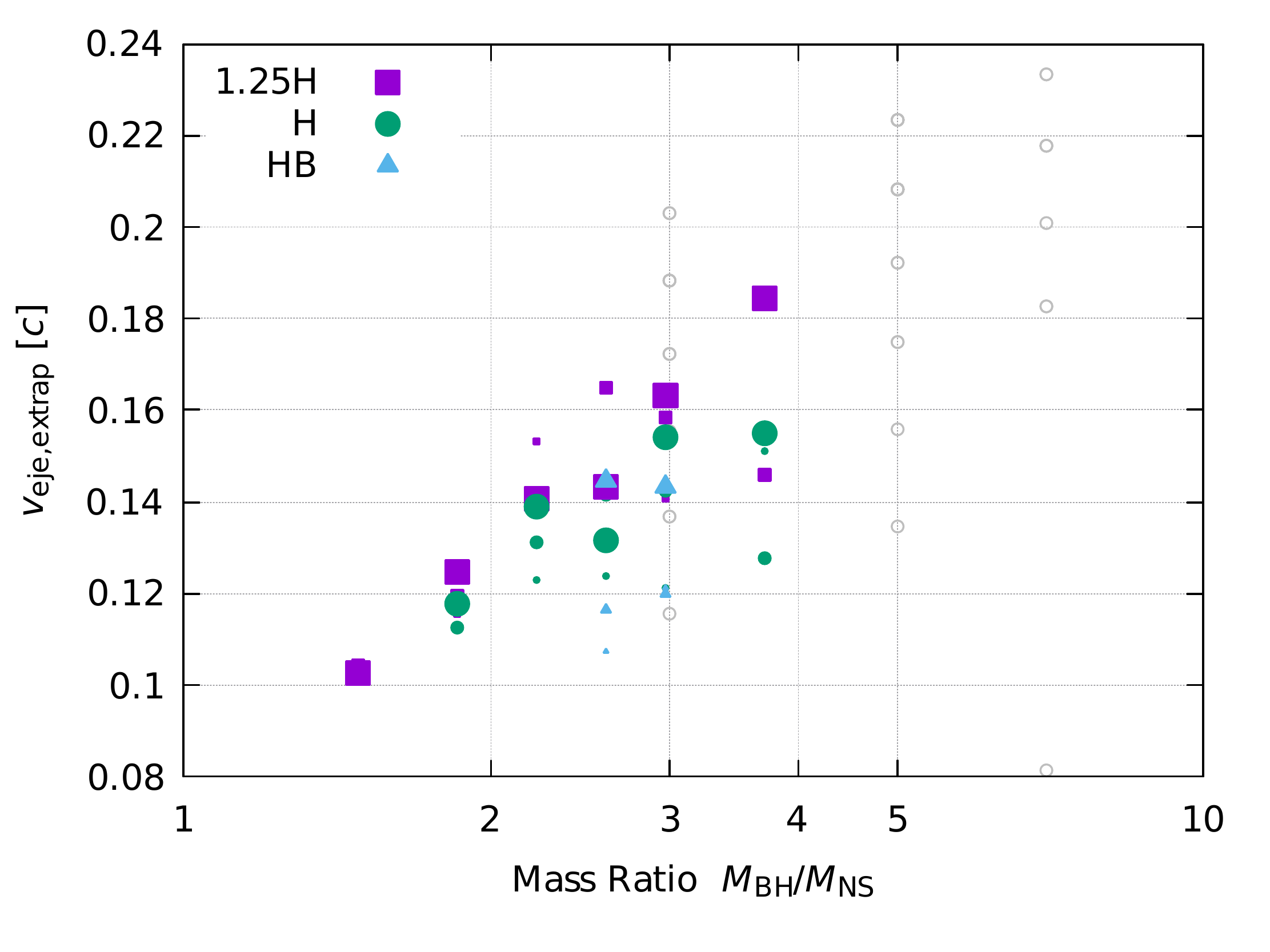}
        \caption{
          The same as Fig.~\ref{fig:rem_er} but for the average velocity of the ejecta extrapolated to $r \to \infty$, $v_{\rm eje,extrap}$,
          as a function of the mass ratio $Q$.
          The results for the model with the ejecta mass larger than $10^{-4} M_{\odot}$ are plotted.
          Data are evaluated at $\SI{12}{\ms}$ after the onset of merger.
          Results from Ref.~\cite{kawaguchi2016jun} are also shown (small open circles).
          Note that data in Ref.~\cite{kawaguchi2016jun} are evaluated at $\SI{10}{\ms}$ after the onset of merger.
        }
        \label{fig:veje_extrap}
      \end{center}
    \end{minipage}

    \\
    \begin{minipage}{0.06\hsize}
      \vspace{5mm}
    \end{minipage}
    \\

  \end{tabular}
\end{figure}

Ejecta velocity is another important quantity that characterizes the ejecta.
Figure \ref{fig:veje_extrap} shows the extrapolated ejecta velocity $v_{\rm eje,extrap}$ as a function of $Q$ for three EOSs.
We also show $v_{\rm eje,extrap}$ for the models simulated in previous studies
(see the small open circles, for which the data are taken from Table 2 of Ref.~\cite{kawaguchi2016jun}).
This figure shows that $v_{\rm eje,extrap}$ tends to decrease as the mass ratio of binary decreases.
By contrast, the dependence on neutron-star EOSs is likely to be weak.
This result is consistent with the results obtained in the merger of binaries consisting of spinning black holes
\cite{kyutoku2015aug,kawaguchi2016jun}.

\subsection{Considerations on the mass ejection} \label{sec:consider_eje}
In Sec.~\ref{sec:eje_mass} we have shown the main results of our simulations focusing on the rest mass and the velocity of the ejecta.
The mass-ratio dependence of the ejecta properties is summarized as follows.
\begin{itemize}
\item
  The ejecta mass increases as the mass ratio decreases for a high mass-ratio regime,
  while it decreases as the mass ratio decreases for a low mass-ratio regime.
  The peak of the ejecta mass is found at $Q\sim3$. 
\item
  The ejecta velocity decreases as the mass ratio decreases. 

\end{itemize}
In this section, we consider the mechanism of mass ejection and
give our interpretation for the mass-ratio dependence of the properties of the ejecta.
~\\

\subsubsection{Matter distribution in the energy-angular momentum phase space}
In order to take a close look at the dynamics of the merger
and explain the mass-ratio dependence of the ejecta mass,  
we analyze the matter distribution in the phase space of specific energy and specific angular momentum.
Figure \ref{fig:H_Q22_EJ} shows the time evolution of the distribution in
the phase space of specific energy $\tilde{E}:=-u_{t}$ and
specific angular momentum $\tilde{J}:=u_{\varphi}=u_y(x-x_{\rm BH})-u_x(y-y_{\rm BH})$
for the models with EOS H.\footnote{
  $\tilde{E}$ and $\tilde{J}$ can be understood as the Killing energy and the Killing angular momentum, respectively,
  if we assume the stationary, axisymmetric  spacetime.}
The matter distribution is obtained by the following integral:

\begin{eqnarray}
  \frac{dM}{d\tilde{E} d\hat{J}}(\tilde{E},\hat{J}) 
  :=
  \lim_{\Delta_1, \Delta_2 \to 0} 
  \frac{1}{\Delta_1 \Delta_2}\int_{\substack{|\tilde{E}-\tilde{E}'|<0.5\Delta_1 \\ |\hat{J}-\hat{J}'|<0.5\Delta_2 }}
  \rho_{*}(\tilde{E}',\hat{J}')d^3x , \label{eq:ang-en} \nonumber \\
\end{eqnarray}
where $\hat{J} := \tilde{J}/M_{\rm BH,f}$. 
Here, $M_{\rm BH,f}$ is the mass of the remnant black hole.
From each panel of Fig.~\ref{fig:H_Q22_EJ}, we can extract the following information for the inspiral, merger, and post-merger stages: 
\begin{description}
\item[1. Inspiral stage] ~\\
  The first panel of Fig.~\ref{fig:H_Q22_EJ} shows the matter distribution in the phase space for the late inspiral stage just prior to the merger.
  The upper limit of the specific energy for each specific angular momentum can be described approximately by $\tilde{E}=\Omega \tilde{J} + C$,
  where $\Omega$ is an orbital angular velocity
  and $C$ is a constant (at each time slice) in the presence of a helical symmetry.\footnote{
    $h\tilde{E}=\Omega h\tilde{J} + C$ is satisfied if we assume the helical symmetry and irrotational fluid \cite{kyutoku2009jun}.
    Here, the helical Killing vector is written as $\xi^{\mu}:=(\partial_{t})^{\mu}+\Omega (\partial_{\varphi})^{\mu}$.
    Therefore, $\tilde{E}=\Omega \tilde{J} + C/h \leq \Omega \tilde{J} + C$ holds.}
  The figure shows that most components have values of $\hat{J}$ smaller than $\hat{J}_{\rm ISCO}$ and will fall into the black hole after the merger. 
\item[2. The onset of merger] ~\\
  The second panel of Fig.~\ref{fig:H_Q22_EJ} shows the matter distribution in the phase space at the onset of merger. 
  This shows that the matter acquires a wide range of specific angular momentum 
  and a fraction of the matter has the specific angular momentum satisfying $\hat{J} \geq \hat{J}_{\rm ISCO}$.
  This can be understood as a result of the angular momentum transport caused by the tidal deformation of the neutron star.
  The associated increase and decrease of the specific energy can also be observed.
  Note that the upper limit of the distribution approximately follows $\tilde{E}=\Omega \tilde{J} + C$ as with the case of the inspiral stage
  but with different values of $\Omega$ and $C$. 
\item[3. $\bm{\approx} \mathbf{1.5 \ ms}$ after the onset of merger] ~\\
  The third panel of Fig.~\ref{fig:H_Q22_EJ} shows the matter distribution in the phase space after the onset of tidal disruption of the neutron star.
  Compared with the second panel, the matter with $\hat{J}<\hat{J}_{\rm ISCO}$ falls into the black hole
  and disappears from the drawing range. 
  The specific energy of matter that remains outside the black hole increases, while the specific angular momentum does not change significantly
  from the onset of merger.
  Therefore, the major effect in this stage is the change in the specific energy by the radial force acting on the matter.
  The matter acquires energy when the large portion of neutron-star matter falls into the black hole
  and the black hole parameters change significantly.
  The time duration for the matter to acquire energy is $0.5$--$\SI{1}{\ms}$.
  We speculate that the instantaneous change in the structure of the spacetime increased the specific energy of a portion of the matter.

  %
  %
  
\item[4. Quasi-steady state after merger] ~\\
  The last panel of Fig.~\ref{fig:H_Q22_EJ} shows the matter distribution in the phase space for a quasi-steady state established after the merger. 
  There are two components observed in this figure.
  One is the component that has non-circular orbits around the black hole (i.e., $E$ has large values for a given value of $\hat{J}$). 
  The phase-space distribution of this component does not change significantly from $\approx \SI{1.5}{\ms}$ after the onset of merger.
  The other is the component that has a circular or nearly circular orbit around the black hole (the components along the magenta dashed curve).
  This constitutes the disk surrounding the black hole.
  Due to fallback and matter interaction in the disk,
  the angular momentum distribution of the latter component is changed significantly from
  $\approx \SI{1.5}{\ms}$ after the onset of merger.

\end{description}

\subsubsection{Model for matter distribution in the specific energy-angular momentum phase space}
\begin{figure*}[]

  \begin{tabular}{cc}
    
    \begin{minipage}[t]{0.50\hsize}
      \begin{center}
        \includegraphics[scale=0.58]{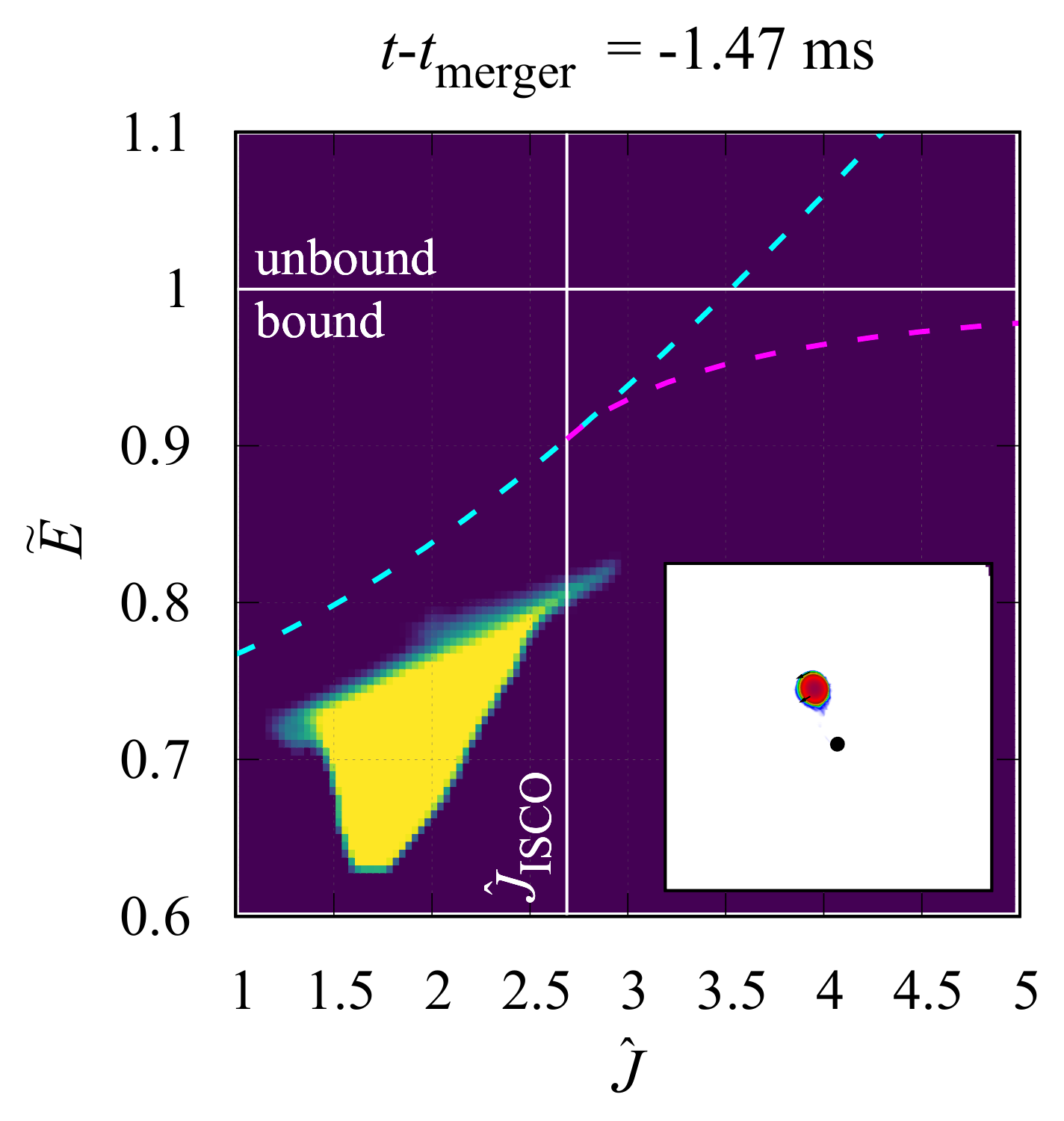}
      \end{center}
    \end{minipage}

    \begin{minipage}[t]{0.50\hsize}
      \begin{center}
        \includegraphics[scale=0.58]{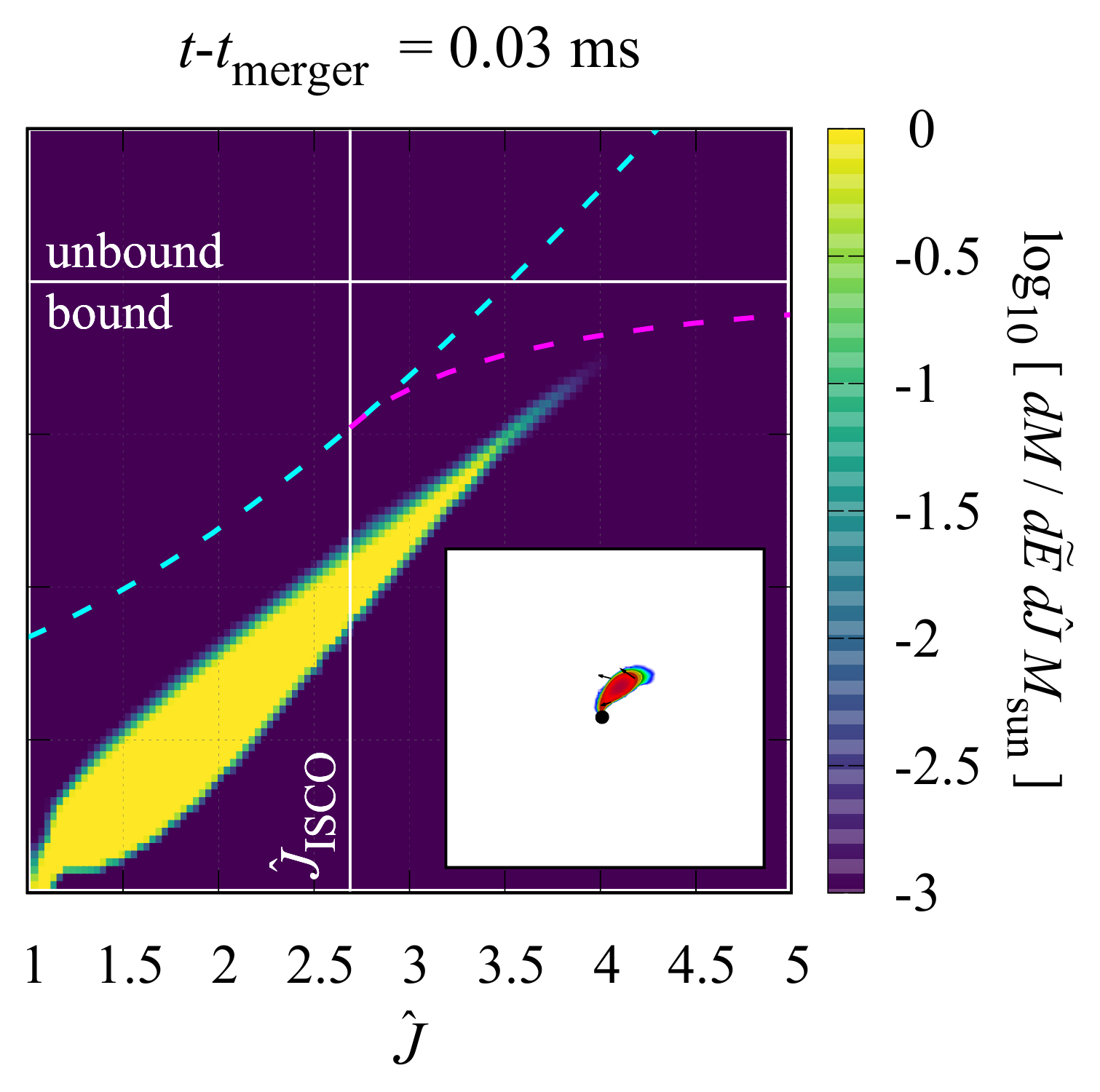}
      \end{center}
    \end{minipage}

  \end{tabular}

  \begin{minipage}{0.06\hsize}
    \vspace{5mm}
  \end{minipage}

  \begin{tabular}{cc}
    
    \begin{minipage}[t]{0.50\hsize}
      \begin{center}
        \includegraphics[scale=0.58]{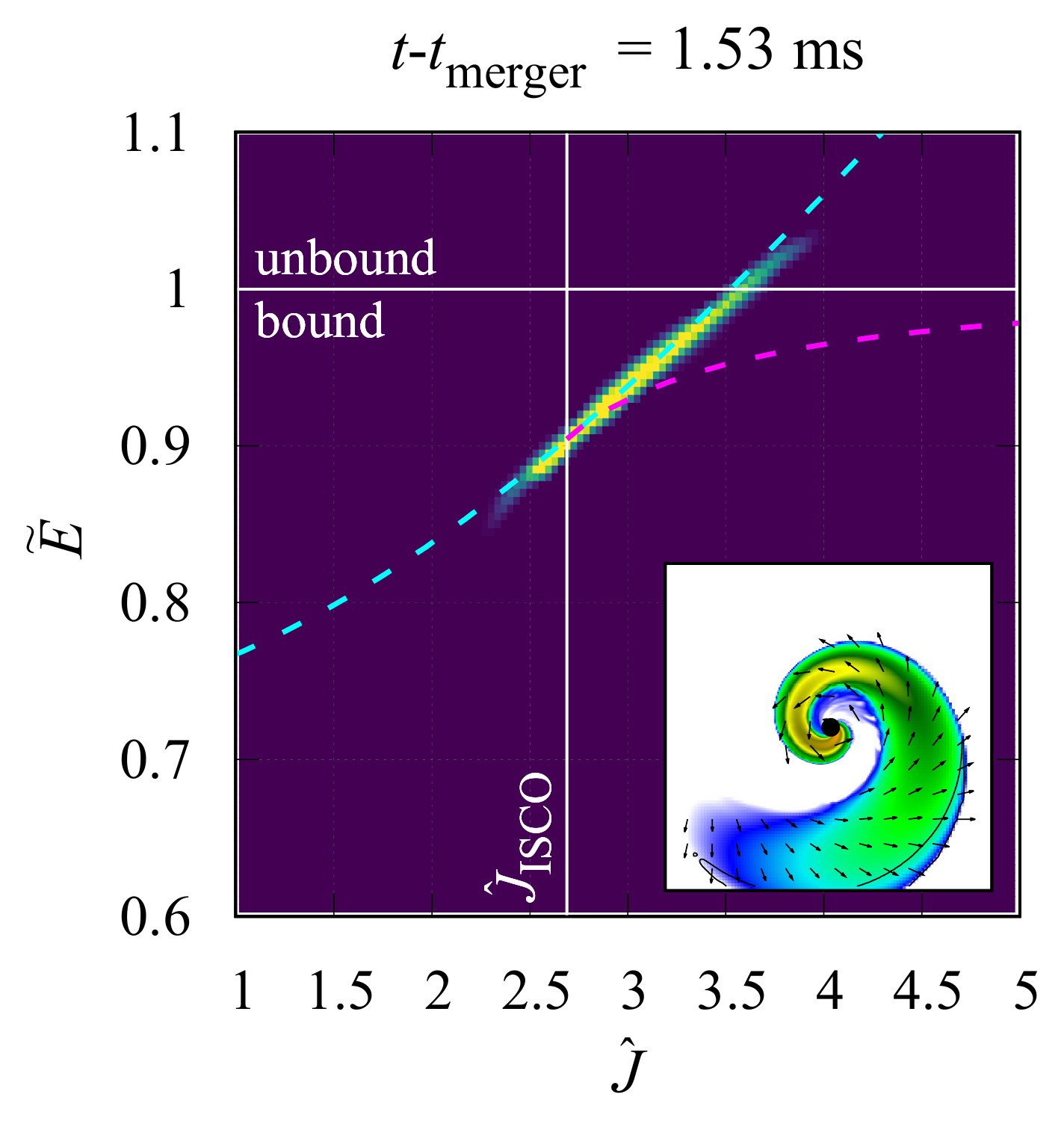}
      \end{center}
    \end{minipage}

    \begin{minipage}[t]{0.50\hsize}
      \begin{center}
        \includegraphics[scale=0.58]{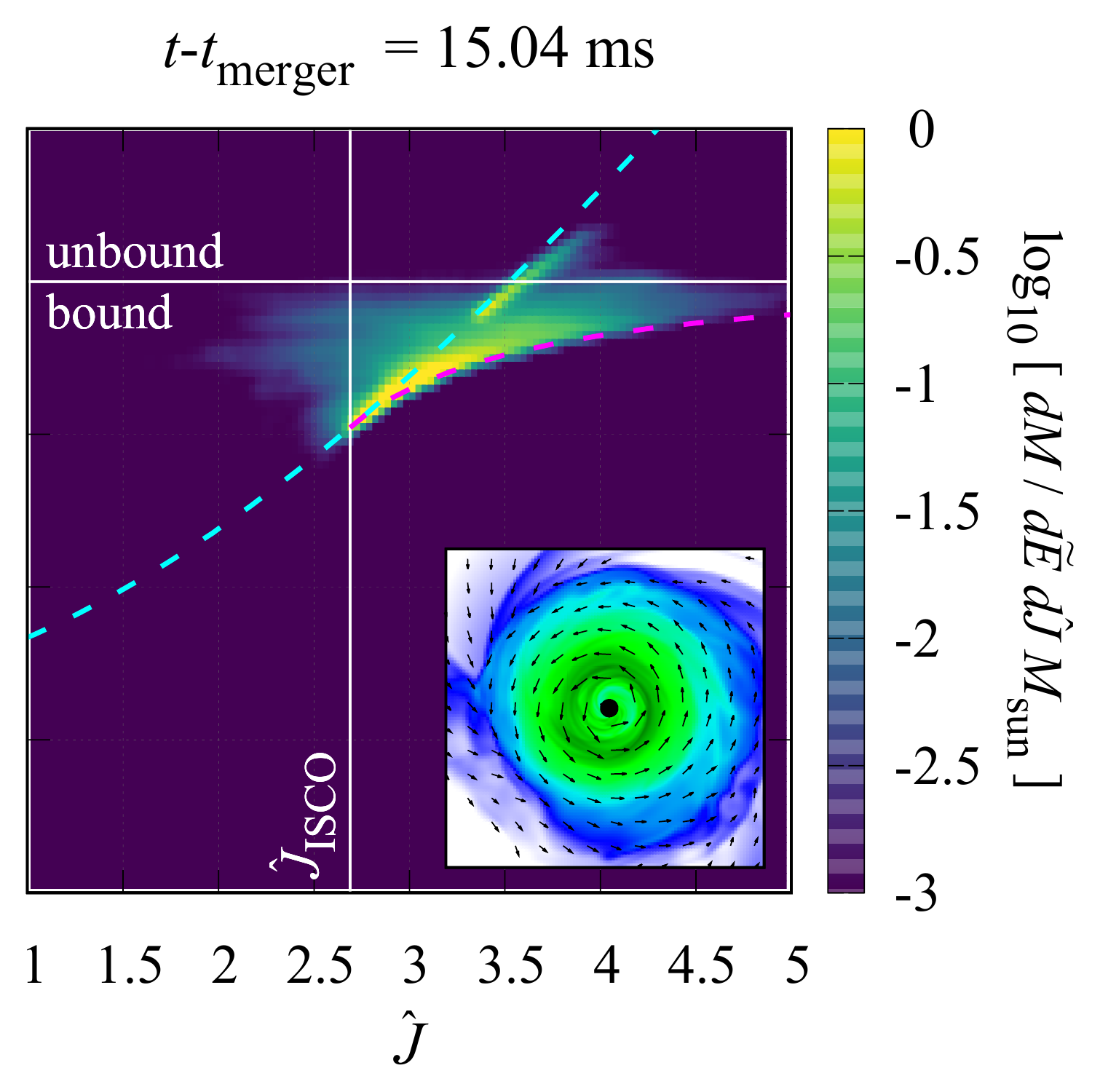}
      \end{center}
    \end{minipage}

  \end{tabular}

  \caption{
    Time evolution of the matter distribution in the phase space of specific energy $\tilde{E}=-u_{t}$ 
    and specific angular momentum normalized by the remnant black-hole mass $\hat{J}=u_{\varphi}/M_{\rm BH,f}$
    at $t-t_{\rm merger} \approx -1.5,\ 0.0,\ 1.5$, and $\SI{15.0}{\ms}$ for the model H\_Q22.
    The contours are obtained by using Eq.~\eqref{eq:ang-en}.
    The cyan dashed curve describes $\tilde{E}=V_{\rm eff}(\hat{J})$ [Eq.~\eqref{eq:kerbh-vpm}].
    This curve shows the effective potential of Kerr spacetime at the orbital angular velocity $\Omega_{\rm merger}$.
    The values of the black-hole mass $M_{\rm BH}$ and the black-hole spin $\chi_{\rm BH}$ used to evaluate Eq.~\eqref{eq:kerbh-vpm}
    are given in Table \ref{tab:rem_bh_p}.
    The magenta dashed curve describes $(\tilde{E},\hat{J})=(\tilde{E}_{\rm sco},\hat{J}_{\rm sco})$ [Eq.~\eqref{eq:kerr-sco}].
    This curve shows the relation of the specific energy and the specific angular momentum
    at stable circular orbits in the Kerr spacetime.
    $\hat{J}_{\rm ISCO}$ is the specific angular momentum at the innermost stable circular orbit around the remnant black hole 
    normalized by its mass.
    The matter with $\hat{J}<\hat{J}_{\rm ISCO}$ falls into the black hole during the merger phase,
    while the matter with $\hat{J}>\hat{J}_{\rm ISCO}$ remains outside the black hole and forms a remnant disk or ejecta.
    The horizontal line of $\tilde{E}=1$ is the boundary of the bound and unbound matter.
    The matter with $\tilde{E}>1$ is unbound from the system and becomes ejecta.
    In each figure, the snapshot of the rest-mass density profile at the corresponding time slice is embedded.
    See Fig.~\ref{fig:H_Q22_xy} for the details of the snapshots.
  }
  \label{fig:H_Q22_EJ}
  
\end{figure*}

In order to deeply understand the numerical results in this paper, 
we here develop a model for the phase-space distribution of the matter after merger using the effective potential of the remnant black hole,
and compare it with the results of numerical simulations in Fig.~\ref{fig:H_Q22_EJ}.
We first consider the phase-space distribution of the matter at $\approx \SI{1.5}{\ms}$ after the onset of merger.
For the analysis, we assume the following:
\begin{itemize}
\item
  The matter motion after merger is determined only by the gravitational effects of the remnant black hole. 
  The mass and the dimensionless spin of the remnant black holes are denoted by $M_{\rm BH,f}$ and $\chi_{\rm BH,f}$, respectively,
  and they are listed in Table \ref{tab:rem_bh_p}.
\item
  All the fluid elements follow the geodesic in the Kerr spacetime starting from the radial position $r_{\rm merger}/M_{\rm BH,f}=((\Omega_{\rm merger}M_{\rm BH,f})^{-1}-\chi_{\rm BH,f})^{2/3}$ in Boyer-Lindquist coordinates,
  with the zero radial velocity, $dr/d\tau=0$, 
  where $\Omega_{\rm merger}$ is the orbital angular velocity at the onset of merger and $\tau$ is the proper time of the fluid elements.
  $\Omega_{\rm merger}$ and $\hat{r}_{\rm merger}:=r_{\rm merger}/M_{\rm BH,f}$ are also listed in Table~\ref{tab:rem_bh_p}.
  This table shows that the value of $\Omega_{\rm merger}$ is primarily determined by the EOS and depends only weakly on the mass ratio.
  
\end{itemize}
Under these assumptions,
the relation between the specific energy and the specific angular momentum is given by
\begin{widetext}
  \begin{eqnarray}
    &&\tilde{E}=V_{\rm eff}(\hat{J})=V_{+}(\hat{r}_{\rm merger},\hat{J}), \nonumber \\
    &&V_{\pm}(\hat{r},\hat{J})
    :=
    \frac{2 \chi_{\rm BH,f} \hat{J} \pm\left(\hat{r}^{2}+{\chi_{\rm BH,f}}^{2}-2 \hat{r}\right)^{\frac{1}{2}}\left[\hat{J}^{2} \hat{r}^{2}
      +\hat{r}\left(\hat{r}^{3}+{\chi_{\rm BH,f}}^{2} \hat{r}+2 {\chi_{\rm BH,f}}^{2}\right)\right]^{\frac{1}{2}}}{\hat{r}^{3}+{\chi_{\rm BH,f}}^{2}(\hat{r}+2 )},
    \label{eq:kerbh-vpm}
  \end{eqnarray}
\end{widetext}
where $\hat{r}:=r/M_{\rm BH,f}$.
We note that $V_{\pm}(\hat{r},\hat{J})$ is obtained from
an effective potential of the geodesic motion around a Kerr black hole \cite{ruffini1971jan}
\begin{widetext}
  \begin{eqnarray}
    \left( \frac{dr}{d\tau}\right)^2&+&V(\hat{r},\tilde{E},\tilde{J})=0, \nonumber \\
    V(\hat{r},\tilde{E},\tilde{J})
    &:=&
    \frac{1}{\hat{r}^{3}}\left[\left\{\hat{r}^{3}+{\chi_{\rm BH,f}}^{2}(\hat{r}+2 )\right\} \tilde{E}^{2}+(2 -\hat{r}) \hat{J}^{2} 
      -4  \chi_{\rm BH,f} \tilde{E} \hat{J}-\hat{r}^{2}(\hat{r}-2 )-{\chi_{\rm BH,f}}^{2} \hat{r} \right] \nonumber \\
    &=&
    \left\{1+\frac{{\chi_{\rm BH,f}}^{2}}{\hat{r}^{3}}(\hat{r}+2) \right\}
    \{\tilde{E}-V_{+}(\hat{r},\hat{J})\} \{\tilde{E}-V_{-}(\hat{r},\hat{J})\} . 
  \end{eqnarray}
\end{widetext}

\begin{figure*}[]
  \begin{tabular}{c}

    \begin{minipage}{0.5\hsize}
      \begin{center}
        \includegraphics[scale=0.45]{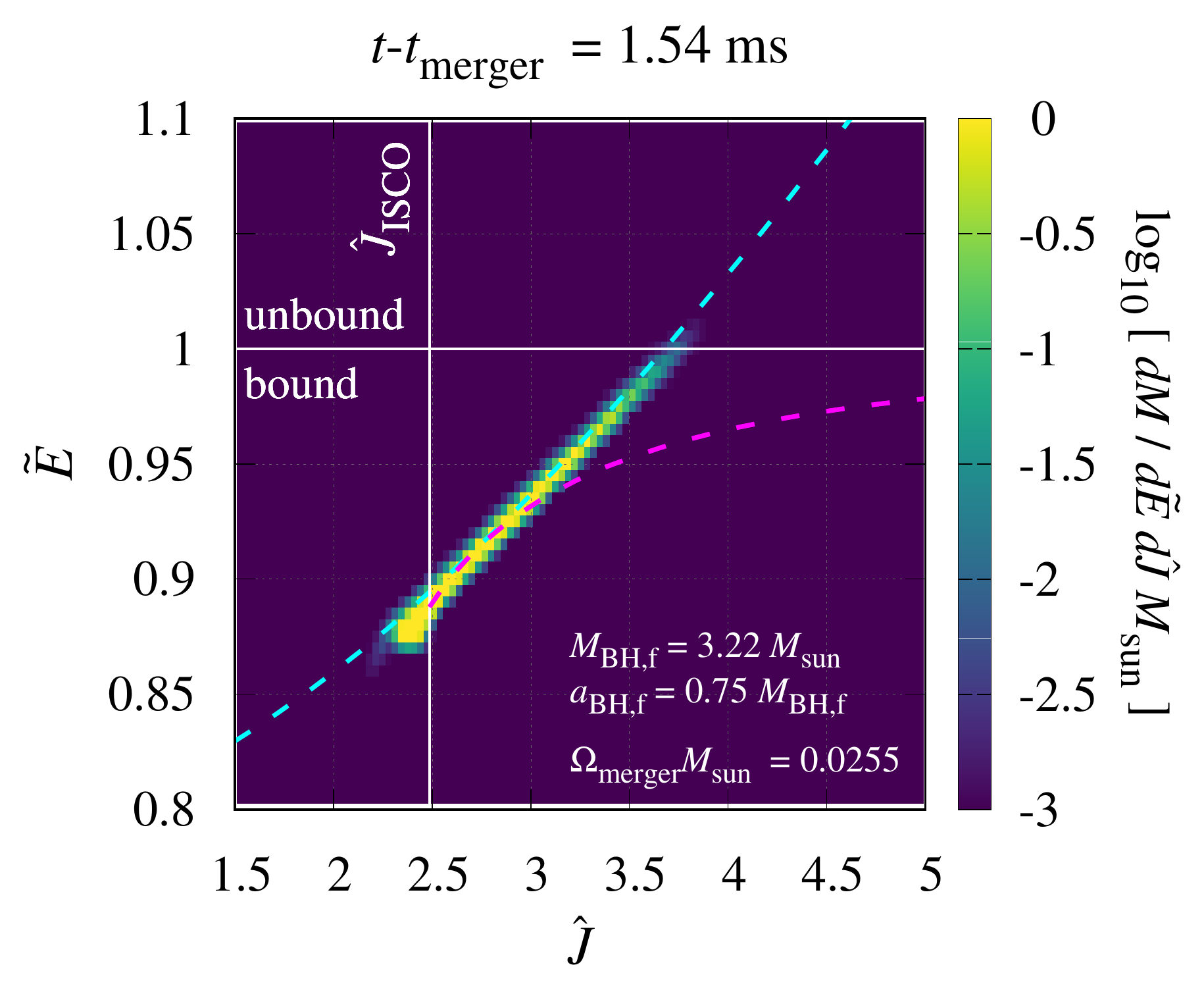}
      \end{center}
    \end{minipage}

    \begin{minipage}{0.5\hsize}
      \begin{center}
        \includegraphics[scale=0.45]{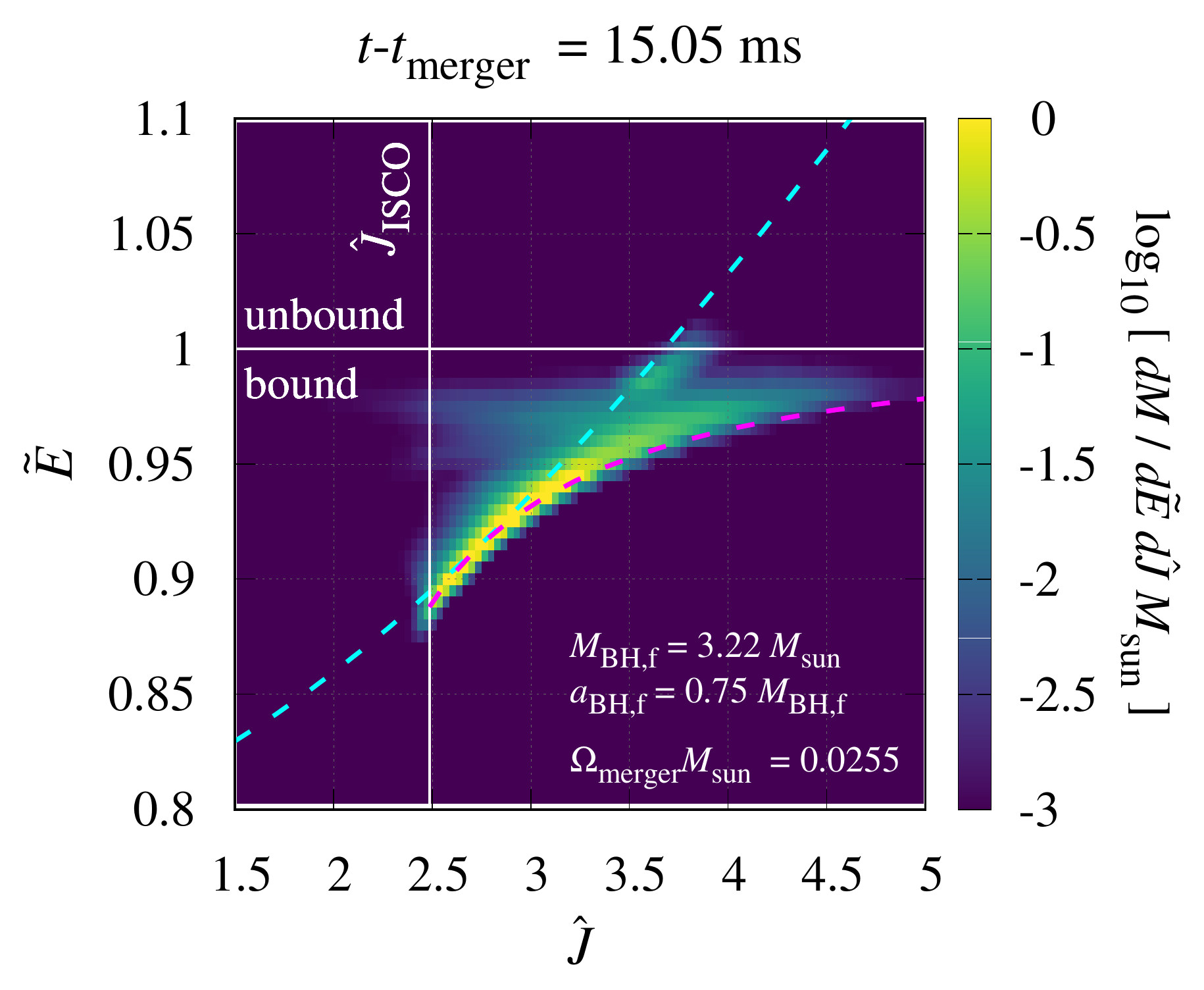}
      \end{center}
    \end{minipage}

  \end{tabular}
  \caption{
    The same as Fig.~\ref{fig:H_Q22_EJ} but for model H\_Q15 at $t-t_{\rm merger} \approx 1.5$, and $\SI{15.0}{\ms}$.
  }
  \label{fig:H_Q15_ea}
  
\end{figure*}

\begin{figure*}[]
  \begin{tabular}{c}

    \begin{minipage}{0.5\hsize}
      \begin{center}
        \includegraphics[scale=0.45]{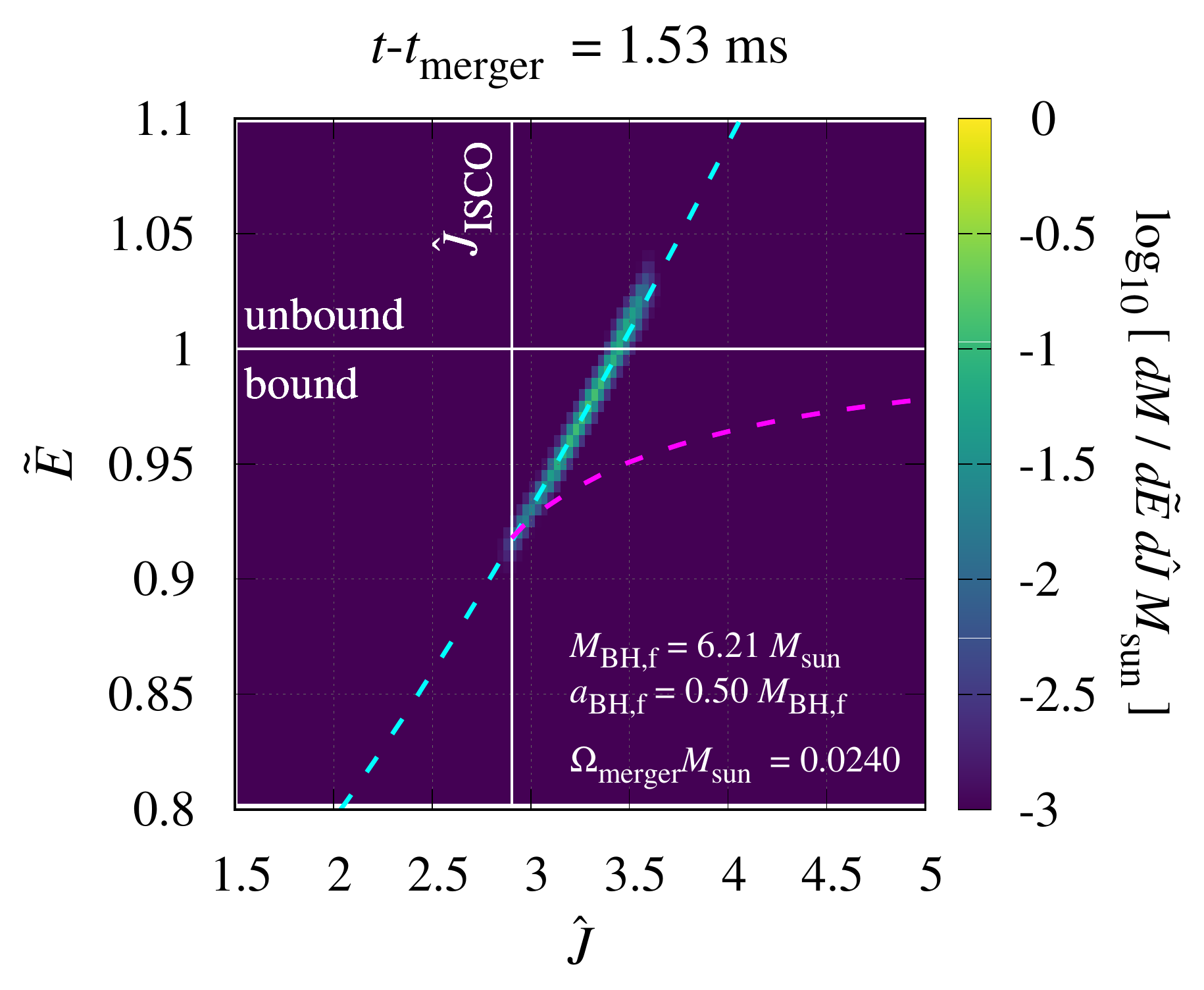}
      \end{center}
    \end{minipage}

    \begin{minipage}{0.5\hsize}
      \begin{center}
        \includegraphics[scale=0.45]{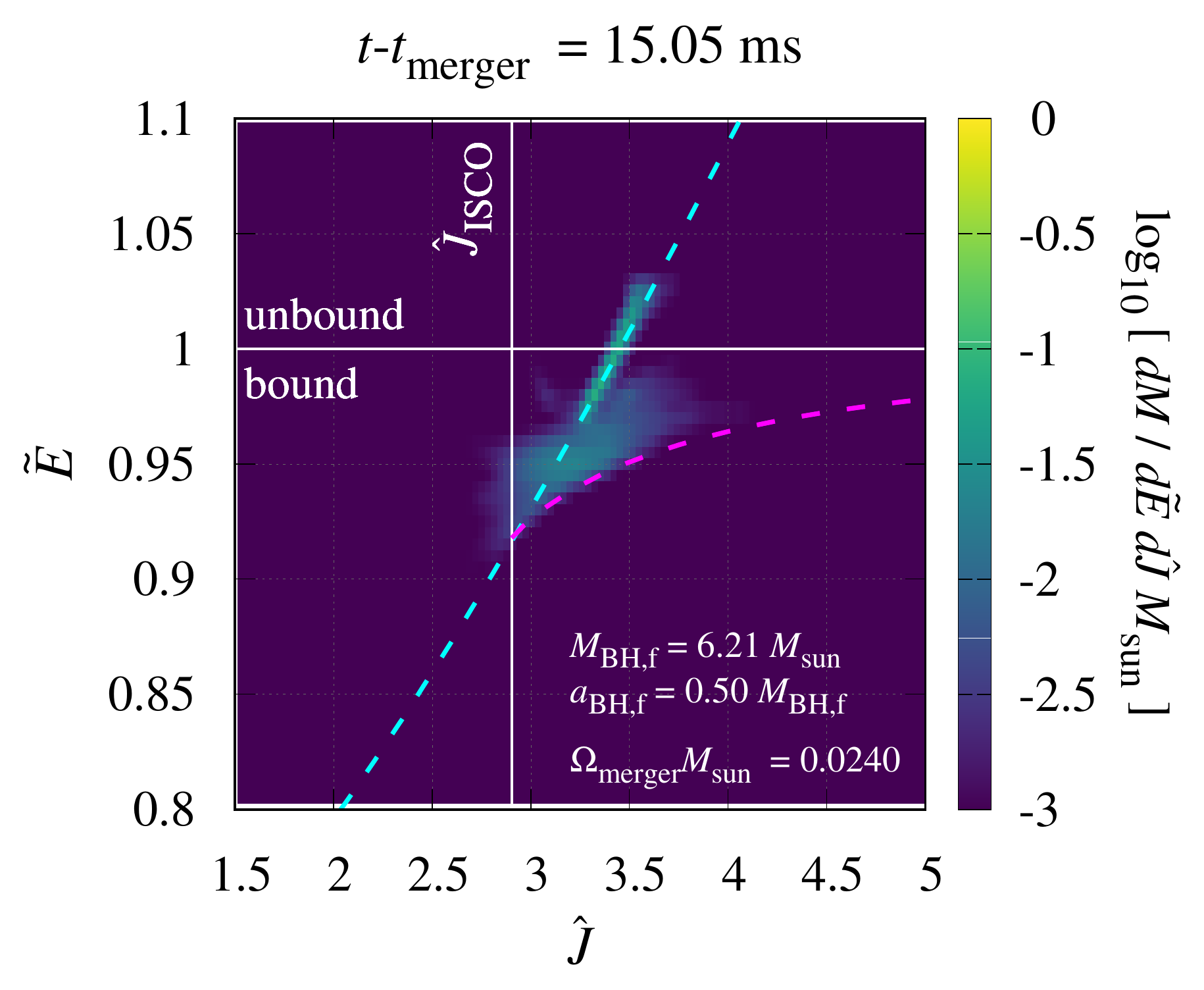}
      \end{center}
    \end{minipage}

  \end{tabular}
  \caption{
    The same as Fig.~\ref{fig:H_Q15_ea} but for model H\_Q37
  }
  \label{fig:H_Q37_ea}
  
\end{figure*}

Equation \eqref{eq:kerbh-vpm} is described by the cyan dashed curve in Fig.~\ref{fig:H_Q22_EJ}.
Comparing the curve describing Eq.~\eqref{eq:kerbh-vpm} and
the phase-space distribution of the matter at $\approx \SI{1.5}{\ms}$ after the onset of merger obtained by the simulation,
we find a reasonable agreement between them in Fig.~\ref{fig:H_Q22_EJ}.
Thus we consider that the model (assumption) given here is consistent with the result of the simulation.
This can also be found in other models:
In Figs.~\ref{fig:H_Q15_ea} and \ref{fig:H_Q37_ea}
we show the same quantity as Fig.~\ref{fig:H_Q22_EJ}
but for the models H\_Q15 and H\_Q37
at $\approx \SI{1.5}{\ms}$ and $\approx \SI{15.0}{\ms}$ after the onset of merger.
For all the models shown in these figures,
the curve describing Eq.~\eqref{eq:kerbh-vpm} is consistent with
the matter distribution at $\approx \SI{1.5}{\ms}$ after the onset of merger obtained by simulations.
All these results validate our model in terms of the effective potential 
for understanding the post-merger phase-space distribution of the matter.

The specific energy-angular momentum distribution of the matter with $\tilde{E}>1$ 
at $\approx \SI{15.0}{\ms}$ after the onset of merger 
approximately agrees with the model given by Eq.~\eqref{eq:kerbh-vpm}.
This implies that the ejecta component moves along the geodesic without any significant matter interaction,
and the conservation of the specific energy and specific angular momentum is approximately satisfied.

We also consider the phase-space distribution of the bound matter in a quasi-steady state after merger.
As found from the velocity distribution of the disk in the fourth panel of Fig.~\ref{fig:H_Q22_xy},
a large portion of the disk matter is in a circular motion.
The relation between specific energy and specific angular momentum
for particles at stable circular orbits is given by \cite{bardeen1972dec}
\begin{eqnarray}
  \tilde{E}_{\rm sco}(\hat{r}) &=& \frac{\hat{r}^{3 / 2}-2 \hat{r}^{1 / 2} + {\chi_{\rm BH,f}} }{\hat{r}^{3 / 4}\left(\hat{r}^{3 / 2}-3  \hat{r}^{1 / 2} + 2 {\chi_{\rm BH,f}} \right)^{1 / 2}}
  \quad (\hat{r} > \hat{r}_{\rm ISCO}) , \nonumber \\
  \hat{J}_{\rm sco}(\hat{r}) &=& \frac{+ \left(\hat{r}^{2} - 2 {\chi_{\rm BH,f}} \hat{r}^{1 / 2}+{\chi_{\rm BH,f}}^{2}\right)}{\hat{r}^{3 / 4}\left(\hat{r}^{3 / 2}-3 \hat{r}^{1 / 2} + 2 {\chi_{\rm BH,f}} \right)^{1 / 2}}
  \quad (\hat{r} > \hat{r}_{\rm ISCO}) , \nonumber \\
\label{eq:kerr-sco}
\end{eqnarray}
where we assumed co-rotating orbits.
Equation \eqref{eq:kerr-sco} is described by the magenta dashed curve in Figs.~\ref{fig:H_Q22_EJ}--\ref{fig:H_Q37_ea}.
The fourth panel of Fig.~\ref{fig:H_Q22_EJ} shows that 
the curve describing Eq.~\eqref{eq:kerr-sco} is consistent with
the phase-space distribution of the disk matter for which the specific energy and specific angular momentum are $\tilde{E} \sim 0.9$--$0.95$ and $\hat{J} \sim 2.5$--$3.5$, respectively.
It shows that the inner region of the disk is supported dominantly by rotation. 
However, there also exists matter with the specific angular momentum appreciably smaller than $\hat{J}_{\rm sco}$ for the specific energy range $\tilde{E} \sim 0.95$--$1.0$.
We consider that the matter in the outer part of the disk has significant pressure support in addition to rotational support.
These disk structures are consistent with that shown in Ref.~\cite{foucart2014jul}.
For other models shown in Figs.~\ref{fig:H_Q15_ea} and \ref{fig:H_Q37_ea},
we also find the rotation-supported component and the component supported by both rotation and pressure
at $\approx \SI{15.0}{ms}$ after the onset of merger. 
Note that for the models H\_Q15 (Fig.~\ref{fig:H_Q15_ea}) and H\_Q22 (Fig.~\ref{fig:H_Q22_EJ}), the former component is the majority,
while for the model H\_Q37 (Fig.~\ref{fig:H_Q37_ea}), the latter component is the majority.

\begin{table}[]
  \centering

  \caption{
    Quantities used in the model given by Eq.~\eqref{eq:kerbh-vpm}.
    $M_{\rm BH,f}$ and $\chi_{\rm BH,f}$ are the mass and the dimensionless spin of the remnant black hole, respectively. 
    $\Omega_{\rm merger}$ is the orbital angular velocity at the onset of merger,
    and $\hat{r}_{\rm merger}$ is the radial position of matter in Boyer-Lindquist coordinates obtained from $\Omega_{\rm merger}$.
  }
  \label{tab:rem_bh_p}
  
  \begingroup
  \setlength{\tabcolsep}{6pt} 
  \renewcommand{\arraystretch}{1.2} 

  \begin{tabular}{c cc cc}
    \hline
    \hline
    Model & $M_{\rm BH,f} [M_{\odot}]$ & $\chi_{\rm BH,f}$ & $\Omega_{\rm merger} [{M_{\odot}}^{-1}]$ & $\hat{r}_{\rm merger}$  \\
    \hline
    125H\_Q15  & 3.20  & 0.75  & $2.24 \times 10^{-2}$ & 5.59 \\ 
    125H\_Q19  & 3.68  & 0.69  & $2.27 \times 10^{-2}$ & 5.02 \\ 
    125H\_Q22  & 4.17  & 0.64  & $2.27 \times 10^{-2}$ & 4.62 \\ 
    125H\_Q26  & 4.67  & 0.59  & $2.30 \times 10^{-2}$ & 4.24 \\ 
    125H\_Q30  & 5.18  & 0.56  & $2.29 \times 10^{-2}$ & 3.95 \\
    125H\_Q37  & 6.21  & 0.50  & $2.24 \times 10^{-2}$ & 3.55 \\ 
    125H\_Q44  & 7.21  & 0.45  & $2.12 \times 10^{-2}$ & 3.34 \\
    \hline
    H\_Q15     & 3.22  & 0.75  & $2.55 \times 10^{-2}$ & 5.08 \\
    H\_Q19     & 3.70  & 0.69  & $2.55 \times 10^{-2}$ & 4.60 \\
    H\_Q22     & 4.20  & 0.64  & $2.55 \times 10^{-2}$ & 4.23 \\
    H\_Q26     & 4.70  & 0.60  & $2.55 \times 10^{-2}$ & 3.91 \\
    H\_Q30     & 5.21  & 0.56  & $2.53 \times 10^{-2}$ & 3.67 \\
    H\_Q37     & 6.21  & 0.50  & $2.40 \times 10^{-2}$ & 3.37 \\
    H\_Q44     & 7.21  & 0.45  & $2.22 \times 10^{-2}$ & 3.22 \\
    \hline
    HB\_Q15    & 3.25  & 0.75  & $2.88 \times 10^{-2}$ & 4.62 \\ 
    HB\_Q19    & 3.73  & 0.69  & $2.89 \times 10^{-2}$ & 4.19 \\ 
    HB\_Q22    & 4.22  & 0.64  & $2.87 \times 10^{-2}$ & 3.87 \\ 
    HB\_Q26    & 4.72  & 0.60  & $2.83 \times 10^{-2}$ & 3.62 \\ 
    HB\_Q30    & 5.22  & 0.56  & $2.77 \times 10^{-2}$ & 3.43 \\
    \hline
    \hline
  \end{tabular}

  \endgroup

\end{table}

%
%
%

%
%
%
%
%

%
%
%
%
%

%
%
%
%
%


\subsubsection{Dependence of the ejecta mass on the mass ratio}
We describe our interpretation for the dependence of the ejecta mass on the mass ratio
using the model for the matter distribution in the phase space and 
the distribution of matter with respect to the specific angular momentum normalized by the mass of the remnant black hole.
Specifically, we compare the results of H\_Q15, H\_Q22, and H\_Q37.

\begin{figure}[]
  \begin{tabular}{c}

    \begin{minipage}{1.0\hsize}
      \begin{center}
        \includegraphics[scale=0.40]{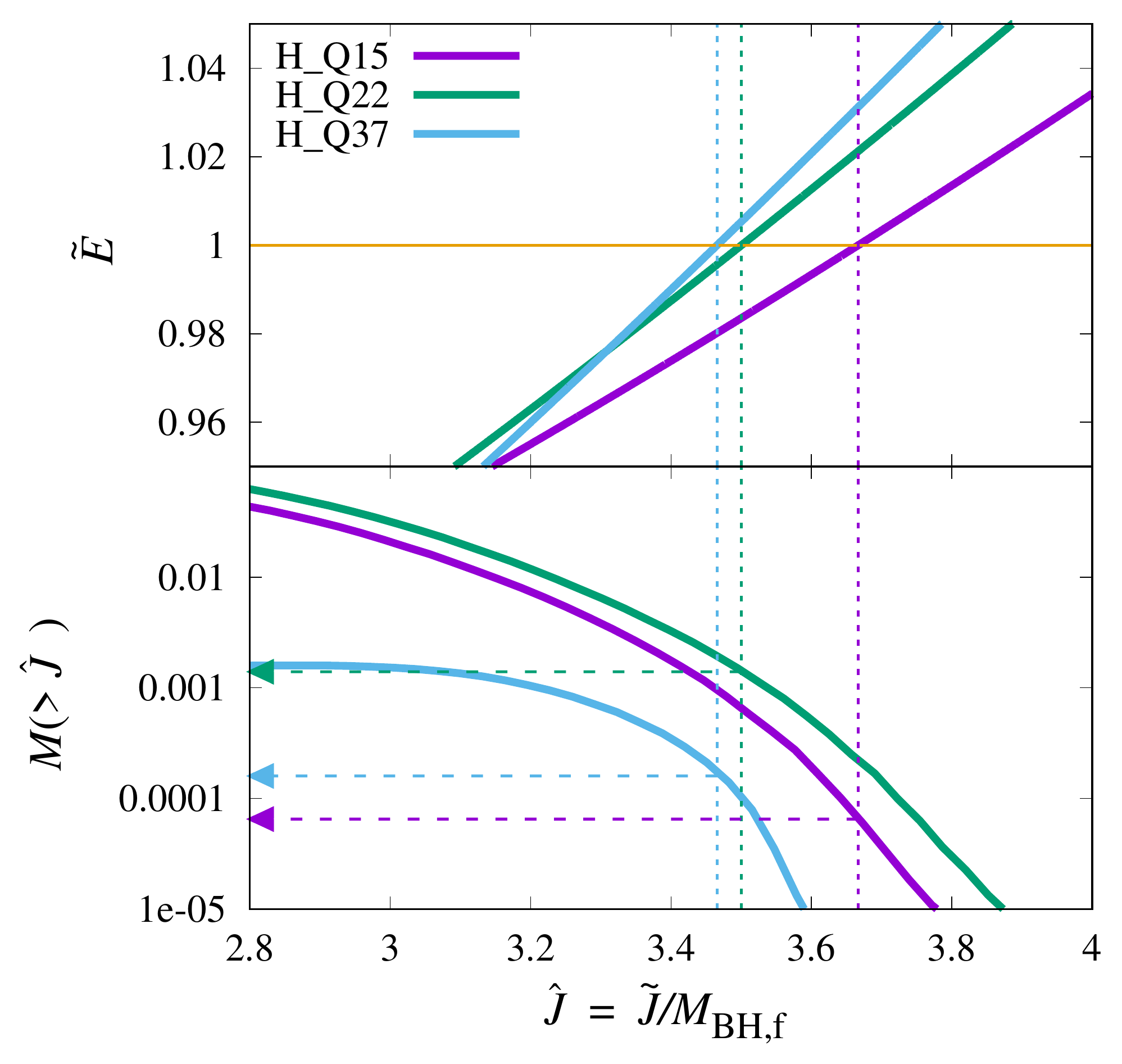}
        \caption{
          The model for the matter distribution in the specific energy-angular momentum phase space given by Eq.~\eqref{eq:kerbh-vpm} (top panel),
          and the specific angular momentum distribution obtained by applying Eq.~\eqref{eq:j_dist} to numerical results of simulations
          at $\approx \SI{1.5}{\ms}$ after the onset of merger (bottom panel).
          The specific angular momentum is always normalized by the mass of the remnant black hole.
          We compare the results for H\_Q15, H\_Q22, and H\_Q37.
          The horizontal line of $\tilde{E}=1$ in the top panel is the boundary of bound and unbound matter.
          The specific angular momentum required for the matter to become ejecta $\hat{J}_{\rm crit}$ is described by the vertical dashed lines.
          As the mass ratio decreases, $\hat{J}_{\rm crit}$ increases.
          The horizontal arrows in the bottom panel show $M(>\hat{J}_{\rm crit})$ for each model.
          The mass-ratio dependence of $M(>\hat{J}_{\rm crit})$ is consistent with the mass-ratio dependence of the ejecta mass.
        }
        \label{fig:meje_m}
      \end{center}
    \end{minipage}

  \end{tabular}
\end{figure}

The top panel of Fig.~\ref{fig:meje_m} shows the mass-ratio dependence of
the specific energy as a function of the specific angular momentum given by Eq.~\eqref{eq:kerbh-vpm}.
This panel shows that as the mass ratio decreases,
the specific angular momentum normalized by the remnant black-hole mass required for the matter to become ejecta (i.e., to achieve $\tilde{E} \geq 1$) increases.
Here, the required specific angular momentum $\hat{J}_{\rm crit}$ is defined as
\begin{eqnarray}
V_{\rm eff}(\hat{J}_{\rm crit})=1.
\label{eq:j_crit}
\end{eqnarray}
$\hat{J}_{\rm crit}$ is also described in Fig.~\ref{fig:meje_m} by the vertical dashed lines.
Here, the mass-ratio dependence of $\hat{r}_{\rm merger}$ is considered to be the main reason for the mass-ratio dependence of $\hat{J}_{\rm crit}$ (see Table \ref{tab:rem_bh_p}). 
The bottom panel of Fig.~\ref{fig:meje_m} shows 
the distribution of matter with respect to the specific angular momentum normalized by the mass of the remnant black hole $M(>\hat{J})$ 
at $\approx \SI{1.5}{\ms}$ after the onset of merger.
Here, the distribution is obtained by the following integral:
\begin{eqnarray}
M(>\hat{J}):=\int_{\hat{J}'>\hat{J}} \rho_{*}(\hat{J}')d^3x .
\label{eq:j_dist}
\end{eqnarray}
By comparing 
$M(>\hat{J})$
for H\_Q15 and H\_Q22, we find that the distribution does not differ significantly.
On the other hand, the values of $M(>\hat{J})$ for H\_Q37 is entirely smaller than that for other two models.
The reason for this is the absence of appreciable tidal disruption for H\_Q37.

The matter with specific angular momentum larger than $\hat{J}_{\rm crit}$ is expected to become ejecta,
and thus the ejecta mass is estimated by $M(>\hat{J}_{\rm crit})$.
$M(>\hat{J}_{\rm crit})$ is shown by the horizontal arrows in the bottom panel of Fig.~\ref{fig:meje_m}.
By comparing the curves of H\_Q15 and H\_Q22 we find that the ejecta mass should decrease as the mass ratio decreases.
Here, the mass-ratio dependence of $\hat{J}_{\rm crit}$ is considered to be the main reason for 
the mass-ratio dependence of the ejecta mass for the low-mass-ratio regime.
On the other hand, by comparing H\_Q22 and H\_Q37 we find that the ejecta mass should increase as the mass ratio decreases. 
In this case, the absence of appreciable tidal disruption for the high-mass-ratio models is the main reason for
the mass-ratio dependence of the ejecta mass.
This mass-ratio dependence of the ejecta mass is consistent with the results presented in Sec.~\ref{sec:eje_mass}.
Quantitatively, the ejecta mass obtained from the analysis in Fig.~\ref{fig:meje_m} agrees with the actual results, $M_{\rm eje}$, within a factor of 2.

\section{Conclusion} \label{sec:conclusion}
We performed numerical relativity simulations for non-spinning low-mass black hole-neutron star mergers with
seven black-hole masses (or mass ratios of the binary) and three neutron-star EOSs.
We considered the cases in which the neutron-star mass is $1.35 M_\odot$.
We paid particular attention to the properties of the matter that remains outside the black hole including ejecta.
We found that the rest mass of such matter is $\sim 0.005$--$0.1 M_{\odot}$ for the case of $Q \alt 3$.
Previous work \cite{shibata2009feb} showed that the rest mass remaining outside the apparent horizon after the merger increases as the mass ratio decreases for $Q \agt 3$.
However, we found that it depends only weakly on the mass ratio for $Q \alt 3$.

Previous work \cite{kyutoku2015aug} also showed that the ejecta mass increases as the mass ratio decreases for $Q \agt 3$ with spinning black holes. 
In this work, we showed that the ejecta mass rapidly decreases as the mass ratio decreases for $Q \alt 3$.
Because it is negligible for $Q \agt 5$ with non-spinning black holes, the ejecta mass shows a peak at the mass ratio $Q \sim 3$.
The peak value of the ejecta mass for models focused on in this work was
$\sim 10^{-3}$--$ 5 \times 10^{-3} M_{\odot}$ depending on the neutron-star EOS.
The average velocity of the ejecta evaluated at $r \to \infty$ is found to be $\sim 0.1$--$0.2c$.
It decreases as the mass ratio decreases, but the dependence on the neutron-star EOSs is weak.

In order to take a close look at the dynamics of the matter in the post-merger stage,
we analyzed the matter distribution in the phase space of the specific energy and the specific angular momentum.
We found that the merger stage can be divided mainly into two stages.
In the first stage, the matter acquires both energy and angular momentum  
as a result of angular momentum transport caused by the tidal deformation of the neutron star.
In the second stage, the matter acquires energy, while there is only a small change in the specific angular momentum.
We do not still understand the physical mechanism for the second stage yet, 
but we speculate that the matter acquires energy due to the instantaneous change in the structure of the spacetime
caused by the infall of the neutron-star matter into the black hole.
As a result of these two stages, a portion of the matter that remains outside the black hole acquires sufficient energy to become ejecta.

The model for the matter distribution in the phase space
suggests that the specific angular momentum normalized by the remnant black-hole mass 
required for the matter to become ejecta increases as the mass ratio of the binary decreases.  
In addition, we find that the distribution of matter with respect to the specific angular momentum normalized by the mass of the remnant black hole
does not depend significantly on the mass ratio for a low-mass-ratio regime.
Combining these two facts, we conclude that the ejecta mass decreases as the mass ratio decreases for the low mass-ratio binaries.

Finally, we list several issues to be explored in the future.
In this paper, we studied the models only with non-spinning black holes 
to focus on the dependence on the mass ratio and the EOS of the neutron star.
However, it is a well-known fact that the black-hole spin influences the merger process significantly \cite{kyutoku2011sep,etienne2009feb}.
We need further studies in order to clarify the parameter dependence of remnant disk and ejecta properties for low-mass black hole-neutron star mergers with a spinning black hole. 
Also, we plan to update the fitting formula for the rest mass remaining outside the apparent horizon after the merger and the ejecta mass
by taking into account the results of this work. 

~\\

\begin{acknowledgments}
  Numerical simulations are performed on
  Cray XC50 at CfCA of National Astronomical Observatory of Japan,
  and XC40 at Yukawa Institute for Theoretical Physics of Kyoto University.
  This work was in part supported by Grant-in-Aid for Scientific Research (Grant Nos. JP16H02183, JP18H01213, JP19K14720, and JP20H00158) of Japanese MEXT/JSPS.
\end{acknowledgments}

~\\

\appendix

\section{Rest mass remaining outside the apparent horizon after the merger for the case of higher mass ratio and spinning black hole} \label{app:rem_hQ_s}
\begin{figure}[h]
  \begin{tabular}{c}

    \begin{minipage}{1.0\hsize}
      \begin{center}
        \includegraphics[scale=0.35]{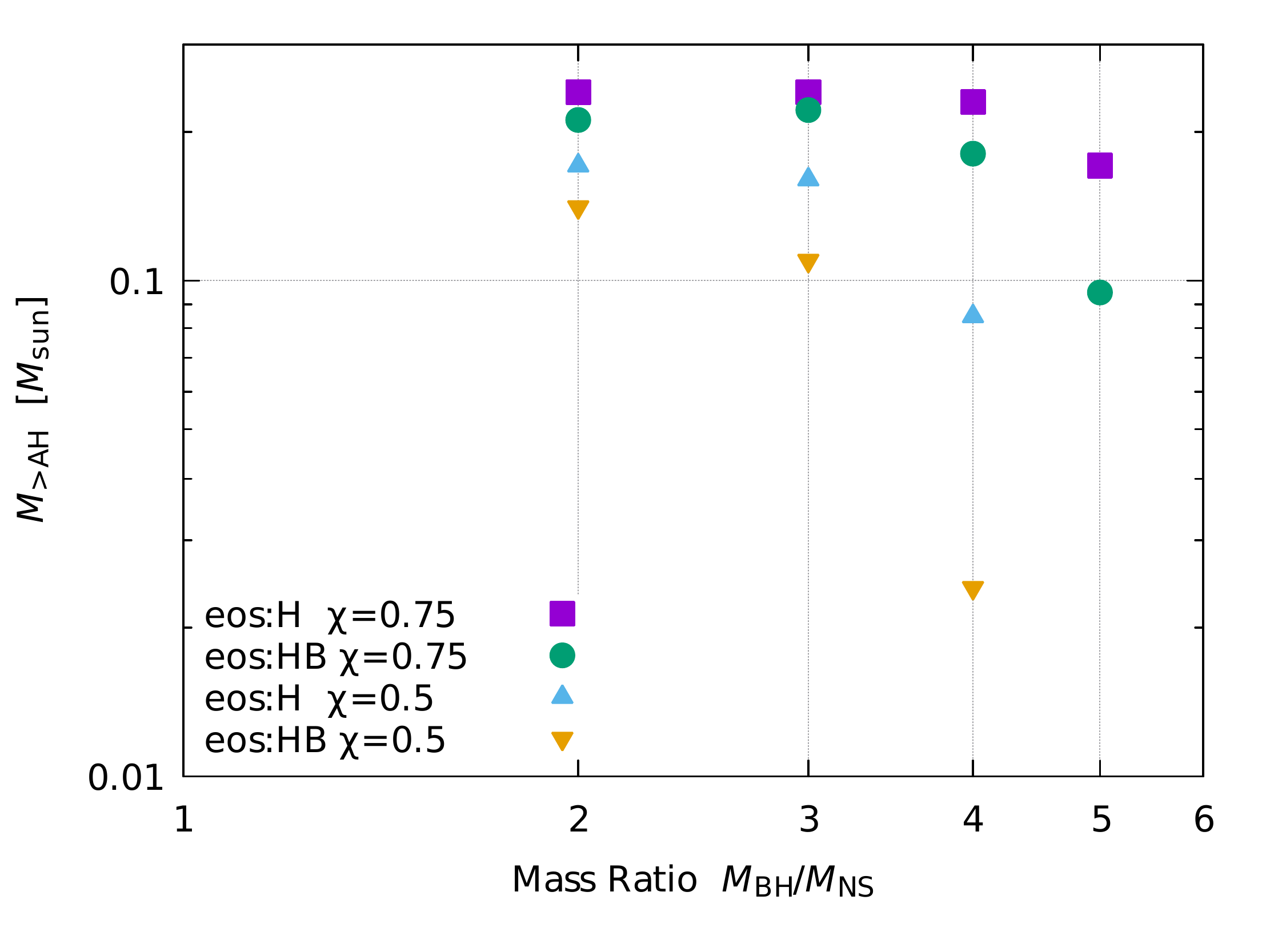}

        \includegraphics[scale=0.35]{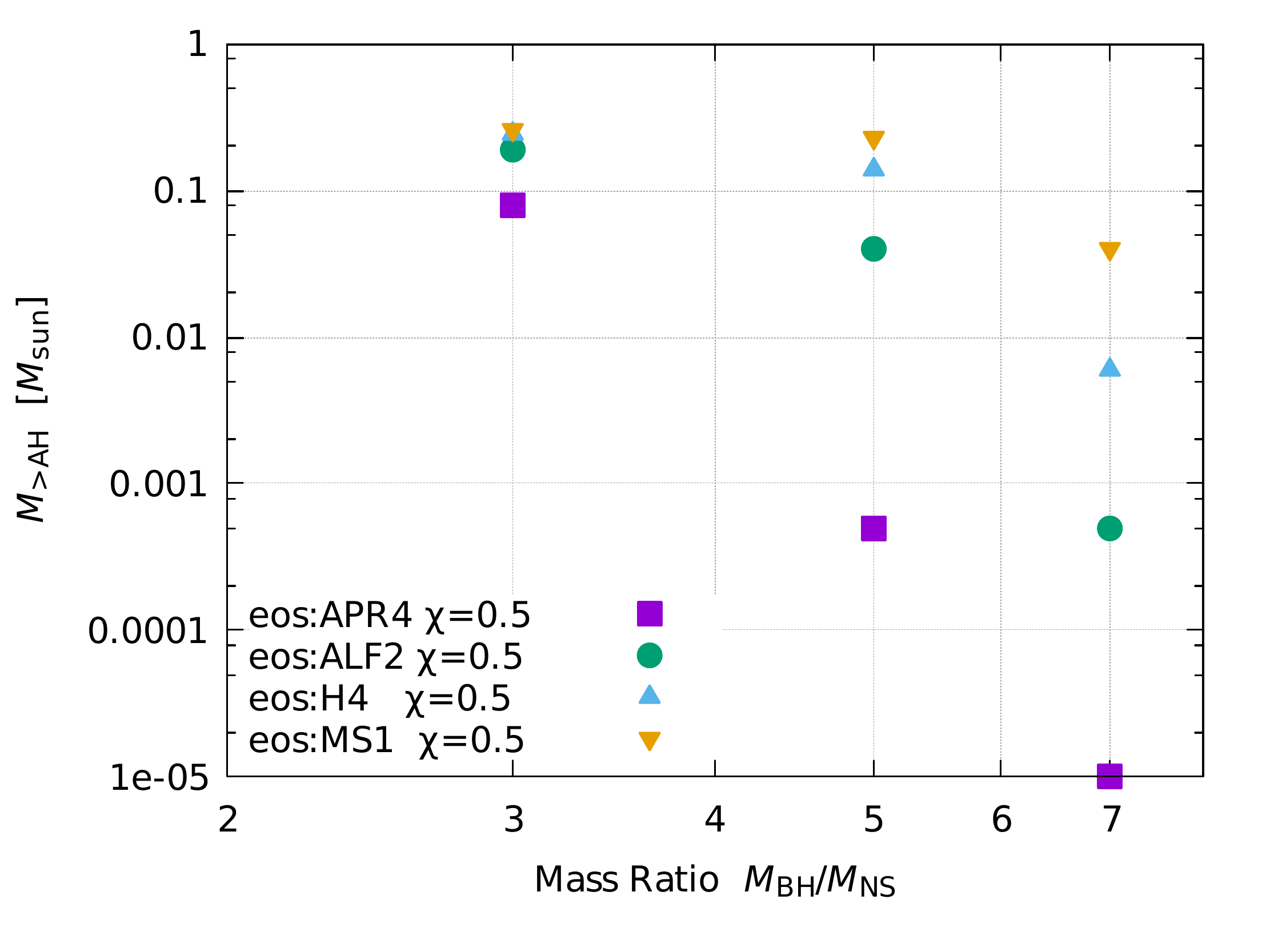}

        \includegraphics[scale=0.35]{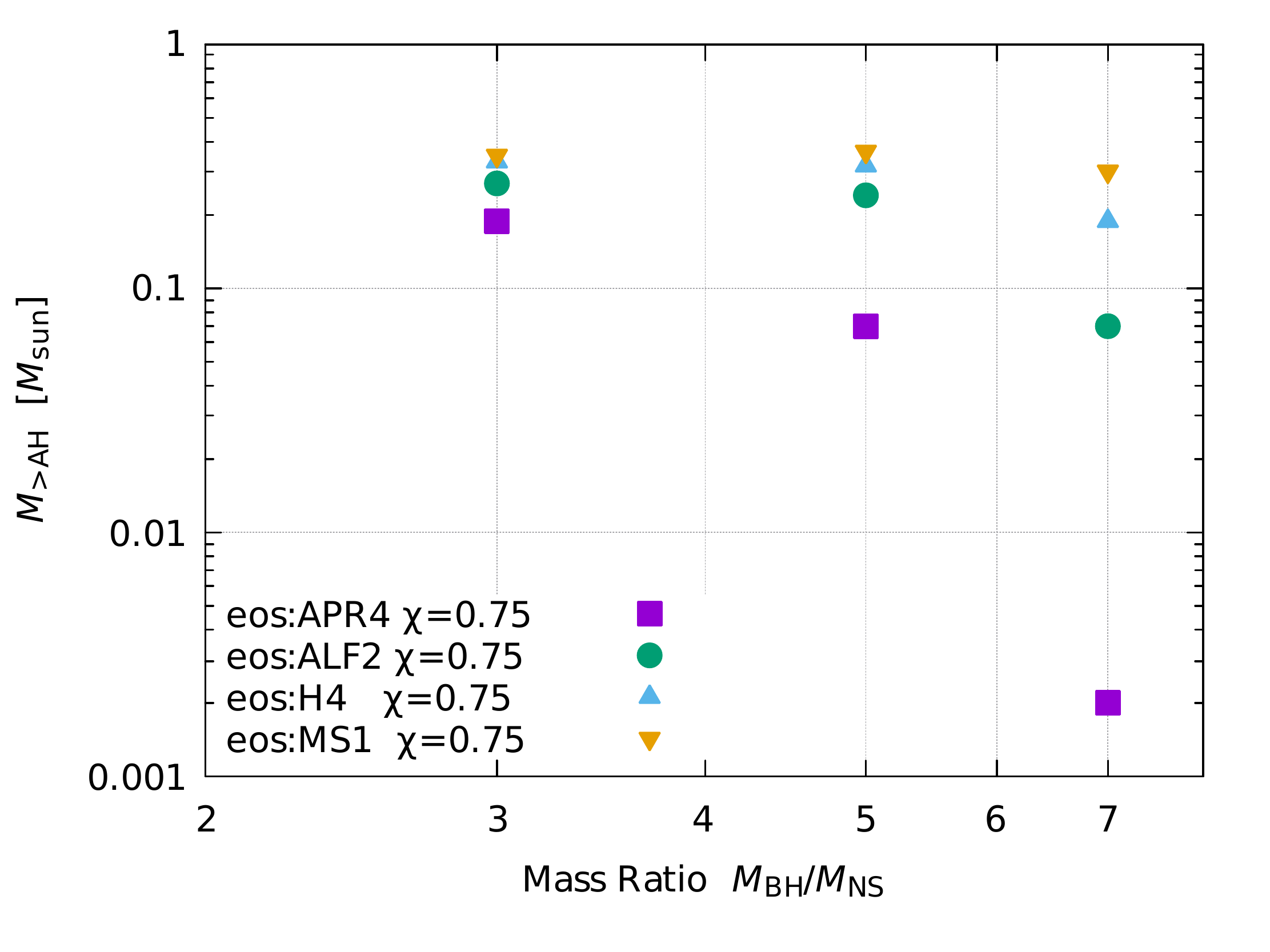}
        
        \caption{
          Dependence of the rest mass remaining outside the apparent horizon after the merger on the mass ratio for binaries with spinning black holes.
          In each panel, EOSs and spin parameters are aligned.
          The data are taken from Ref.~\cite{kyutoku2011sep} (top panel) and Ref.~\cite{kyutoku2015aug} (middle and bottom panels).
          The compactness of the neutron star with a mass $1.35 M_{\odot}$ is
          $0.138,0.147,0.161$, and $0.180$ for EOSs MS1, H4, ALF2, and APR4, respectively. 
        }
        \label{fig:rem_kyu}
      \end{center}
    \end{minipage}

  \end{tabular}
\end{figure}

Though it was not pointed out clearly,
the weak mass-ratio dependence of the rest mass remaining outside the apparent horizon after the merger in a low mass-ratio regime
can be seen for the merger of binaries consisting of spinning black holes
from the results of previous simulations \cite{kyutoku2011sep,kyutoku2015aug}.
We reanalyzed the results of Refs.~\cite{kyutoku2011sep,kyutoku2015aug} focusing on the mass-ratio dependence. 
Figure \ref{fig:rem_kyu} shows that the mass-ratio dependence of the rest mass located outside the apparent horizon tends to be weaker as the mass ratio decreases
for models with higher black-hole spins and stiffer neutron-star EOSs.
It should not be off base to expect the same tendency
for models with lower black-hole spins and softer neutron-star EOSs
in the mass-ratio regime lower than the scope of Refs.~\cite{kyutoku2011sep,kyutoku2015aug}.
We expect that this is a universal behavior irrespective of the black-hole spin.
It should also be noted that the mass ratio at which the rest mass remaining outside the apparent horizon after the merger starts leveling off depends on
the compactness of the neutron star (or neutron-star EOS) and the black-hole spin.

\bibliography{paper}

\end{document}